\documentclass[manuscript]{aastex}

\usepackage{graphicx,amsmath,booktabs,threeparttable,url}

\usepackage[breaklinks]{hyperref}
\usepackage[hyphenbreaks]{breakurl}

\usepackage{CJKutf8}
\usepackage{threeparttable}
\usepackage[usenames,dvips]{color}
\usepackage[normalem]{ulem}
\usepackage{subfigure}

\newcommand{\cntext}[1]{\begin{CJK*}{UTF8}{bsmi}#1\end{CJK*}}

\usepackage{natbib}

\begin{document}

\title{SIMULATIONS OF THE SYMBIOTIC RECURRENT NOVA V407~CYG. I.~ACCRETION AND SHOCK EVOLUTIONS}

\author{Kuo-Chuan Pan (\cntext{潘國全})$^{1}$, Paul M. Ricker$^2$, and Ronald E. Taam$^{3,4}$}
\affil{$^1$Physik Department, Universit\"{a}t Basel, Klingelbergstrasse 82, CH-4056 Basel, Switzerland; kuo-chuan.pan@unibas.ch}
\affil{$^2$Department of Astronomy, University of Illinois at Urbana$-$Champaign, 1002 West Green Street, Urbana, IL 61801, USA; pmricker@illinois.edu}
\affil{$^3$Department of Physics and Astronomy, Northwestern University, 2145 Sheridan Road, Evanston, IL 60208, USA; r-taam@northwestern.edu}
\affil{$^4$Academia Sinica Institute of Astronomy and Astrophysics, P.O. Box 23-141, Taipei 10617, Taiwan; taam@asiaa.sinica.edu.tw}

%-------------------------------------------------------------------------------------------------------------
%  Abstract    *** 250 words limit *** ApJ
%--------------------------------------------------------------------------------------------------------------

\begin{abstract}

The shock interaction and evolution of nova ejecta with a wind from a red giant star
in a symbiotic binary system are investigated via three-dimensional hydrodynamics simulations. 
We specifically model the March 2010 outburst of the symbiotic recurrent nova V407~Cygni
from the quiescent phase to its eruption phase. The circumstellar density enhancement 
due to wind-white dwarf interaction is studied in detail.  
It is found that the density-enhancement 
efficiency depends on the ratio of the orbital speed to the red giant wind speed.  Unlike 
another recurrent nova, RS~Ophiuchi, we do not observe a strong disk-like density enhancement,  
but instead observe an aspherical density distribution with $\sim 20\%$ higher density 
in the equatorial plane than at the poles.  To model the 2010 outburst, we consider several 
physical parameters, including the red giant mass loss rate, nova eruption energy, and ejecta mass.  
A detailed study of the shock interaction and evolution reveals that the interaction 
of shocks with the red giant wind generates strong Rayleigh-Taylor instabilities. In 
addition, the presence of the companion and circumstellar density enhancement greatly alter 
the shock evolution during the nova phase. The ejecta speed after sweeping out most of the circumstellar 
medium decreases to $\sim 100-300$~km~s$^{-1}$, depending on model, which is consistent with 
the observed extended redward emission in [N~II] lines in April 2011. 

\end{abstract}


%-------------------------------------------------------------------------------------------------------------
%  Introduction
%--------------------------------------------------------------------------------------------------------------

\section{INTRODUCTION}

The symbiotic recurrent nova (SyRN) V407~Cyg consists of a white dwarf (WD) and a Mira-type red 
giant (RG, \citealt{1990MNRAS.242..653M}).  In March 2010, the gamma-ray emission from its outburst 
was detected by {\it Fermi}-LAT (Large Area Telescope, \citealt{2009ApJ...697.1071A}), 
showing the first evidence for a gamma-ray nova \citep{2010Sci...329..817A}.  Another gamma-ray 
SyRN, V745~Sco, was also detected by {\it Fermi}-LAT in February 2014 \citep{2014ATel.5879....1C}.
Since the RGs in symbiotic binaries could produce a high-density circumstellar medium (CSM) through 
their stellar wind, gamma-ray emission in SyRNe could originate in high-energy particle 
acceleration that happens when blast waves pass through the high-density CSM 
\citep{2010Sci...329..817A, 2013A&A...551A..37M}. 

Recently, classical novae (CNe) have also been classified as a new type of gamma-ray source 
\citep{2014Sci...345..554A}.  However, the lack of a high-density CSM in CNe cannot explain 
the origin of gamma-ray emission as in SyRNe, requiring a new mechanism for gamma-ray CNe.  
\cite{2014Natur.514..339C} studied the high-resolution radio maps of the gamma-ray CN V959~Mon, 
proposing that the fast binary motion in CNe could shape the nova ejecta by transferring its angular momentum. 
Therefore, high-energy particles could be accelerated at the interface between equatorial and polar regions.   

Since particle acceleration in CNe and SyRNe is highly associated with shock propagation 
in their nova eruptions, these new discoveries motivate us to study the shock evolution in nova systems. 
Additionally, SyRNe have long been considered as Type Ia supernova (SN~Ia) progenitors 
\citep{2001ApJ...558..323H, 2012BASI...40..393K}, although the nature of the progenitor systems 
of SNe~Ia remains uncertain \citep{2014ARA&A..52..107M}.  However, evidence of interaction between 
SN Ia ejecta with a CSM has been found in some SNe~Ia in late-type spiral galaxies 
\citep{2007Sci...317..924P, 2012Sci...337..942D, 2013ApJS..207....3S}, suggesting that at least some 
SNe~Ia could originate in symbiotic binaries. 

The gamma-ray emission from the 2010 outburst of V407~Cyg has been studied by \cite{2013A&A...551A..37M} 
through a semi-analytical study of diffusive shock acceleration and non-thermal emission in 
V407~Cyg.  They found that the observed gamma-ray light curve could be fitted by requiring a 
circumbinary density enhancement (CDE) around the WD.  Multi-dimensional hydrodynamics 
simulations by \cite{2012MNRAS.419.2329O} further support the need for a CDE around the WD.
They simulated a blast wave passing through a dense CSM in V407~Cyg including a sophisticated 
treatment of thermal conduction and radiative cooling \citep{2012MNRAS.419.2329O}, finding that 
the observed X-ray light curves could be reproduced if a CDE exists around the WD.  However, they 
do not include the asymmetric effect of binary motion, and the distribution of CDE is artificially 
imposed in their simulations without modeling the WD-wind interaction in the quiescent phase.  

On the other hand, \cite{2008A&A...484L...9W} performed three-dimensional hydrodynamics simulations of 
another SyRN, RS~Oph, from the quiescent phase to a nova eruption.  In their simulations, they found 
a disk-like CDE concentrated in the orbital plane, and their shock evolution is consistent 
with the observations of the 2006 nova outburst \citep{2006Natur.442..276S, 2006ApJ...652..629B}. 
Although some similarity between RS~Oph and V407~Cyg has been found in their eruptions' spectra 
\citep{2011MNRAS.410L..52M, 2011A&A...527A..98S}, the physical parameters of the binary systems are 
quite different \citep{2013AASP....3...75D}, especially for their orbital periods (separations).  The 
orbital period in RS~Oph is much shorter ($P_{\rm orb} \sim 1.25$~yr) than in V407~Cyg ($P_{\rm orb} 
\sim 43$~yr).  Therefore, the amount and distribution of density enhancement could be different in these two systems.

In this paper, we investigate the evolution of the SyRN V407~Cyg from the quiescent phase to a nova 
eruption.  In the next section, we review and summarize some observational facts about the binary 
system of V407~Cyg and its previous outbursts.  The numerical methods and the initial setup are presented 
in Section~\ref{sec_methods}.  Our simulation results with different red giant wind loss rates and 
nova eruption parameters are reported and discussed in Section~\ref{sec_results}.  In the final 
section, we summarize our results and conclude. 

%-------------------------------------------------------------------------------------------------------------
% Observations
%--------------------------------------------------------------------------------------------------------------
\section{OBSERVATIONS OF V407~CYG}

As mentioned in the above section, the SyRN V407~Cyg consists of a massive WD and a Mira-type M6~III RG 
\citep{1990MNRAS.242..653M}.  The Mira-type RG has a pulsation period of $763$~day \citep{2003ARep...47..777K}. 
From the dust obscuration and the long-term optical light curve, an orbital period of $\sim 43$~yr 
was inferred by \cite{1990MNRAS.242..653M}, corresponding to an orbital separation of $16$~AU for a 
$1M_\odot$ RG and a $1.2 M_\odot$ WD.  The distance of V407~Cyg is still unclear. Some distance estimates
using the period-luminosity relation give a range from $1.7$~kpc \citep{2003ARep...47..777K} to $2.7$~kpc 
\citep{1990MNRAS.242..653M}.  However, \cite{2012ApJ...761..173C} suggest a possible distance $\gtrsim 3.0$~kpc
using the interstellar Na~I D absorption lines, since the period-luminosity relation is not well established 
for the long pulsation period of V407~Cyg \citep{2008MNRAS.386..313W} (see a more detailed argument 
in \citealt{2012ApJ...761..173C}).

V407~Cyg has been monitored in the optical for more than seven decades. Three nova eruptions were 
reported in 1936, 1998, and 2010 \citep{2010ApJS..187..275S}, although some undetected eruptions could 
have happened in this interval. The March 2010 eruption is the most interesting 
due to the use of modern astronomical instrumentation and the detection of gamma-rays. 
From the measurement of an $H_\alpha$ emission line-width, an ejecta mass $M_{\rm ej} \sim 10^{-6} M_\odot$, 
eruption energy $E_{\rm ej} \sim 10^{44}$~ergs, and ejecta speed $v_{\rm ej} \sim 3,200 \pm 345$~km~s$^{-1}$ 
could be estimated \citep{2010Sci...329..817A, 2011ApJS..197...31S}.  An ejecta mass of $M_{\rm ej} \sim 10^{-6} 
M_\odot$ could also reproduce the observed velocity evolution, but may underestimate the RG mass loss rate 
\citep{2012ApJ...761..173C}.

In general, the observed mass loss rate $\dot{M}_{\rm wind}$ of a Mira giant spans a wide range, 
varying from $\sim 10^{-8} M_\odot$~yr$^{-1}$ to $\sim 10^{-4} M_\odot$~yr$^{-1}$ with an average 
$\langle\dot{M}_{\rm wind}\rangle \sim 6 \times 10^{-7} M_\odot$~yr$^{-1}$ \citep{1985ApJ...292..640K}.
The radio light curve from the 2010 outburst of V407~Cyg suggests a mass loss rate $\dot{M}_{\rm wind} 
\sim 2 \times 10^{-6} M_\odot$~yr$^{-1}$ (for $v_{\rm wind} = 10$~km~s$^{-1}$ and $d=3$~kpc; 
\citealt{2012ApJ...761..173C}), but this mass loss rate is one order of magnitude higher than the 
estimated value from the X-ray light curve ($\dot{M}_{\rm wind} \sim 10^{-8} - 10^{-7} M_\odot$~yr$^{-1}$, 
\citealt{2012ApJ...748...43N, 2013A&A...551A..37M}).  To obtain better agreement between the X-ray and 
radio estimates, a binary separation wider than 16~AU may be required \citep{2012ApJ...761..173C}.

The wind speed of a giant star correlates with the star's luminosity \citep{2010A&A...523A..18D, 
2011ApJ...741....5P}.  Taking account of the observed luminosity $L\sim 10^4 L_\odot$ \citep{2011MNRAS.410L..52M} 
and the uncertainty of the distance of V407~Cyg, a RG wind speed of $10-20$~km~s$^{-1}$ can be estimated. 
In addition, a wind speed of $\sim 10$~km~s$^{-1}$ could be also estimated 
from the optical P~Cygni line profiles \citep{2010Sci...329..817A}.
However, a higher wind speed of $30$~km~s$^{-1}$ is estimated from the equivalent width of the Na~I 
D lines in the quiescent phase \citep{2003ARep...47..889T} 
and in the eruption phase \citep{2011A&A...527A..98S, 2012ApJ...748...43N}.

%-------------------------------------------------------------------------------------------------------------
%  Numerical Methods
%--------------------------------------------------------------------------------------------------------------
\section{NUMERICAL METHODS \label{sec_methods}}

\subsection{Simulation Code}
The simulation code used in this study is FLASH\footnote{\url{http://flash.uchicago.edu}}~version~4 
\citep{2000ApJS..131..273F, 2008PhST..132a4046D, 2013JCoPh.243..269L}.  FLASH is a grid-based, parallel, 
multidimensional hydrodynamics code based on block-structured adaptive mesh refinement (AMR).  The split 
piecewise parabolic method (PPM) solver \citep{1984JCoPh..54..174C} is used to solve for the motion of 
compressible fluids.  We use active particle clouds to represent the WD and the core of the RG 
companion.  Particle clouds are utilized instead of single particles to avoid the force anisotropy 
problem in cloud-in-cell interpolation \citep{2008ApJ...672L..41R, 2012ApJ...750..151P}.  Gravitational 
forces between particle clouds and gas and self-gravity are solved using a multigrid Poisson solver 
\citep{2008ApJS..176..293R}.  
Nuclear burning is ignored in our simulations, since it takes place in the envelope of the 
white dwarf, which is not resolved. 
We also assume that magnetic fields do not affect the shock evolution.
 
\subsection{Initial Setup}
To model V407~Cyg and its 2010 eruption, we consider a three-dimensional (3D) simulation box with a size 
of 400~AU~$\times$~400~AU~$\times$~400~AU.  Initially, a $1M_\odot$ RG is located at the center (origin) of 
the simulation box, and a $1.2 M_\odot$ WD is located on the positive $x-$axis with a binary separation of 16~AU, 
corresponding to an orbital period of about 43~yr. The orbital plane is set on the $x-y$ plane and the 
direction of angular momentum is set to be the positive $z-$axis.  Particle clouds representing 
the WD and the RG have a radius of $10^{13}$~cm, which is slightly larger than three smallest zone spacings 
($\Delta x_{\rm min} = 2.93 \times 10^{12}$~cm). We use  9 levels of refinement based on the magnitudes of the 
second derivatives of gas density and pressure.  Each AMR block contains $8^3$ zones in the 3D simulation box, 
corresponding to an effective uniform resolution of $2048^3$.  To save computation time, we reduce the maximum 
AMR level based on the distance to the center of mass, giving an effective resolution of $\Delta x^i \sim 0.61 
\times d$ at the $i$th level of AMR, where $d$ is the distance to the center of mass of the WD-RG binary.
The highest resolution region contains the central 65~AU and the first forced AMR decrement occurs at 
$d \sim 32$~AU.  The second AMR decrement occurs at $d \sim 63$~AU, the third at $d \sim 127$~AU, and so on. 
Outflow boundary conditions for fluids and isolated boundary conditions for the Poisson solver are used. 

We assume the RG has a radius of $2.2$~AU, a constant wind speed $v_{\rm wind} = 20$~km~s$^{-1}$ in the 
co-rotating frame, and a constant mass loss rate $\dot{M}_{\rm wind}$.  Therefore, the initial gas density 
$\rho$ is distributed using  Equation~\ref{eq_rho}:
\begin{equation}
\rho(r) = \frac{\dot{M}_{\rm wind}}{4\pi r^2 v_{\rm wind}}  \label{eq_rho} = 1.12 \times 10^{-14} 
\left( \frac{\dot{M}_{\rm wind}}{10^{-6}~{\rm M_\odot~yr^{-1}}} \right)
\left( \frac{v_{\rm wind}}{20~{\rm km~s^{-1}}} \right)^{-1}
\left( \frac{r}{1~{\rm AU}} \right)^{-2}~{\rm g~cm^{-3}},
\end{equation}
where $r$ is the distance to the RG.  We assume a gamma-law equation of state with $\gamma=1.1$ and wind 
temperature $T_{\rm wind}=7,000$~K to mimic the thermodynamic behavior of gas based on photoionization 
models of symbiotic binary systems \citep{1987A&A...182...51N, 1993A&A...278..209N}.

We perform quiescent-phase simulations up to $\sim 2.0 - 2.5$ orbital periods, since the density distribution 
becomes quasi-steady in the co-rotating frame after about $1.5$ orbital periods.  During 
the quiescent phase, we reset the RG wind using Equation~\ref{eq_rho} for the region with $r < 5$~AU
to maintain the RG wind. 

To model a SyRN eruption, we artificially impose a Sedov-like explosion on the location of the WD.  The 
eruption has an ejecta mass $M_{\rm ej}$, eruption energy $E_{\rm ej}$, and averaged ejecta speed $v_{\rm ej}$. 
To avoid grid effects, the eruption covers a small spherical region with a radius equal to 
fifteen smallest zone spacings \citep{2012ApJ...750..151P}.  Within the erupting region, a uniform 
density distribution and linear velocity profile in radius are used.  We assume the kinetic energy 
is three-fourths of the total eruption energy \citep{1996ApJ...471..279D}.
The mass and kinetic energy of the CDE within the exploding region 
and the orbital speed of the WD are added to the eruption. 
Once the SyRN eruption is imposed, we reduce the resetting radius for the RG wind to the radius of the RG, 
and assume a completely absorbing surface on the RG.  In reality, this assumption is not precisely 
correct, and a reverse shock should be generated when the ejecta reach the surface of the RG. 
We ignore these effects since the purpose of the paper is not to focus on the response of the RG, 
but on the global shock evolution.

%-------------------------------------------------------------------------------------------------------------
%  Results
%--------------------------------------------------------------------------------------------------------------
\section{RESULTS AND DISCUSSION \label{sec_results}}

We perform runs with several different sets of values for the RG wind and SyRN eruption parameters, 
such as  $\dot{M}_{\rm wind}, E_{\rm ej}$, and $M_{\rm ej}$ (see Table~\ref{tab_simulations}), and present our 
simulation results in this section.  In the first subsection, the evolution of wind focused by the WD during 
the quiescent phase is described.  The asymmetry due to WD-wind interaction 
is investigated as well.  In Section~\ref{sec_eruption}, we study the shock evolution dependence on 
different RG wind loss rates and different eruption energies and ejecta masses.

\subsection{Quiescent Phase}

We perform two quiescent-phase simulations by varying the mass loss rate from 
$\dot{M}_{\rm wind}= 10^{-6} M_\odot$~yr$^{-1}$ (model~M6) to 
$\dot{M}_{\rm wind}= 10^{-7} M_\odot$~yr$^{-1}$ (model~M7; see Table:~\ref{tab_simulations}) 
to study the effects on the CDE.
The quiescent phase evolution of both models is qualitatively the same,
but the gas density distributions scale with their mass loss rates.
In this subsection, we take model~M6 as the reference simulation 
and describe the quiescent-phase evolution in detail. 

\begin{figure}
\begin{center}
\epsscale{1.0}
\plotone{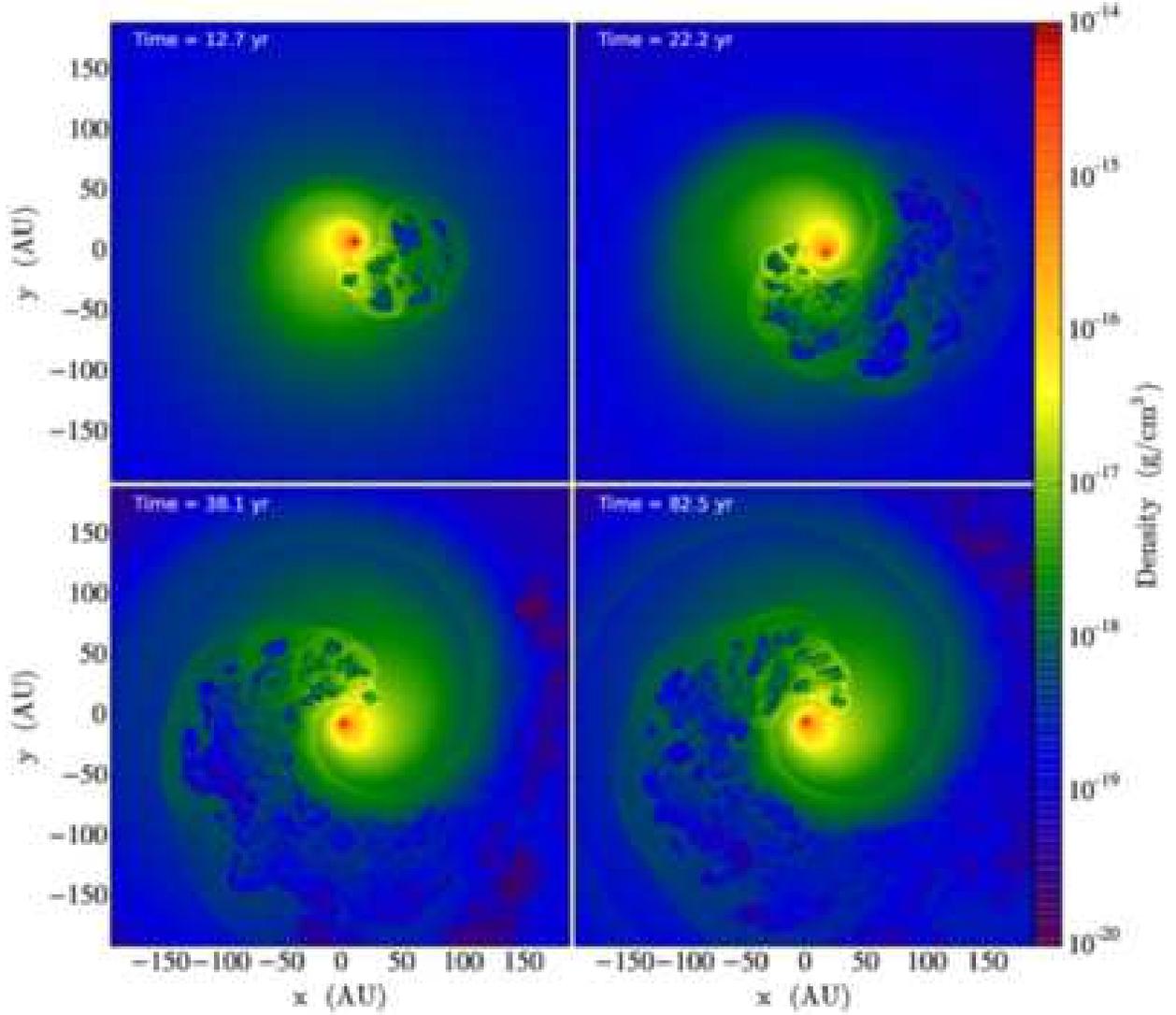}
\end{center}
\caption{\label{fig_M6_dens_x-y} 
Gas density distribution in the orbital plane for model~M6 at different labeled times. 
The color scale indicates the logarithm of the gas density in g~cm$^{-3}$.}
\end{figure}

\begin{figure}
\begin{center}
\epsscale{1.0}
\plotone{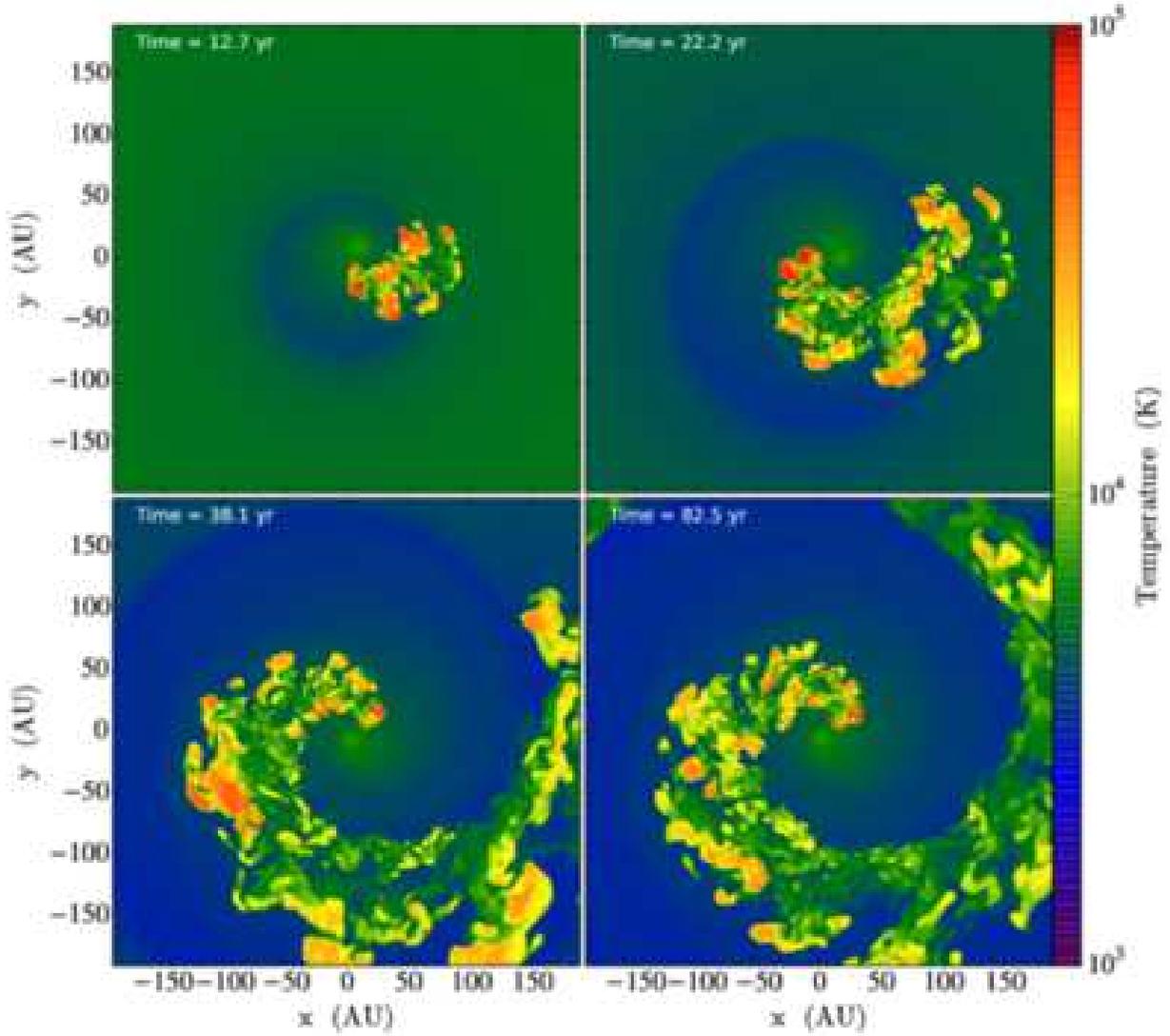}
\end{center}
\caption{\label{fig_M6_temp_x-y} 
Gas temperature distribution in the orbital plane for model~M6 at different labeled times. 
The color scale indicates the logarithm of the gas temperature in $K$.}
\end{figure}

Figure~\ref{fig_M6_dens_x-y} demonstrates a typical evolution of the gas density 
in the orbital plane during the quiescent phase (model~M6).
Since the initial wind density distribution (see Equation~\ref{eq_rho}) neglects the binary interaction, 
a turbulent and hot spiral arm quickly forms when the RG wind interacts with the WD.
This turbulent spiral arm takes about an orbital period ($\sim 40$~yr) to reach a quasi-steady 
large-scale state in the co-rotating frame. 
We also show the corresponding gas temperature distribution in Figure~\ref{fig_M6_temp_x-y}.

\begin{figure}
\begin{center}
\epsscale{0.8}
\plotone{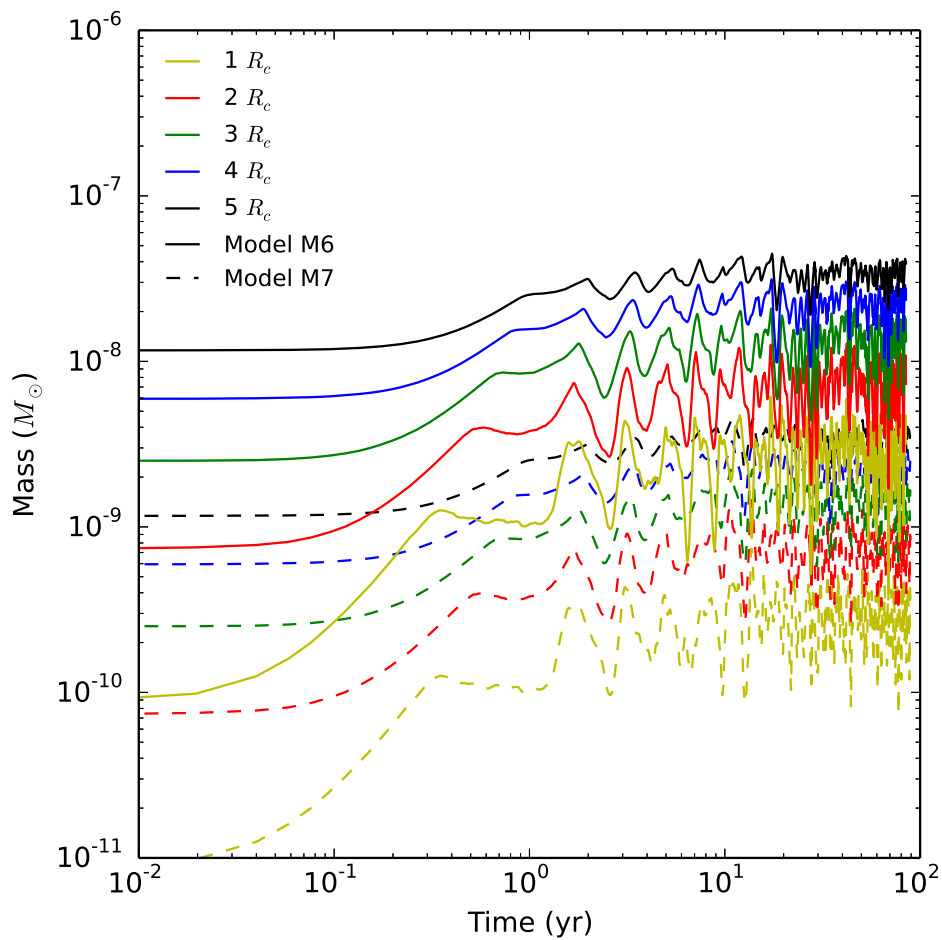}
\end{center}
\caption{\label{fig_accretions} Mass in the vicinity of the WD companion as determined for spherical 
control surfaces of different radii ($R_c = 10^{13}$~cm) as a function of time during the quiescent phase. 
Different line styles indicate different mass loss rates of the RG wind (see Table~\ref{tab_simulations}). }
\end{figure}

\begin{figure}
\begin{center}
\epsscale{0.8}
\plotone{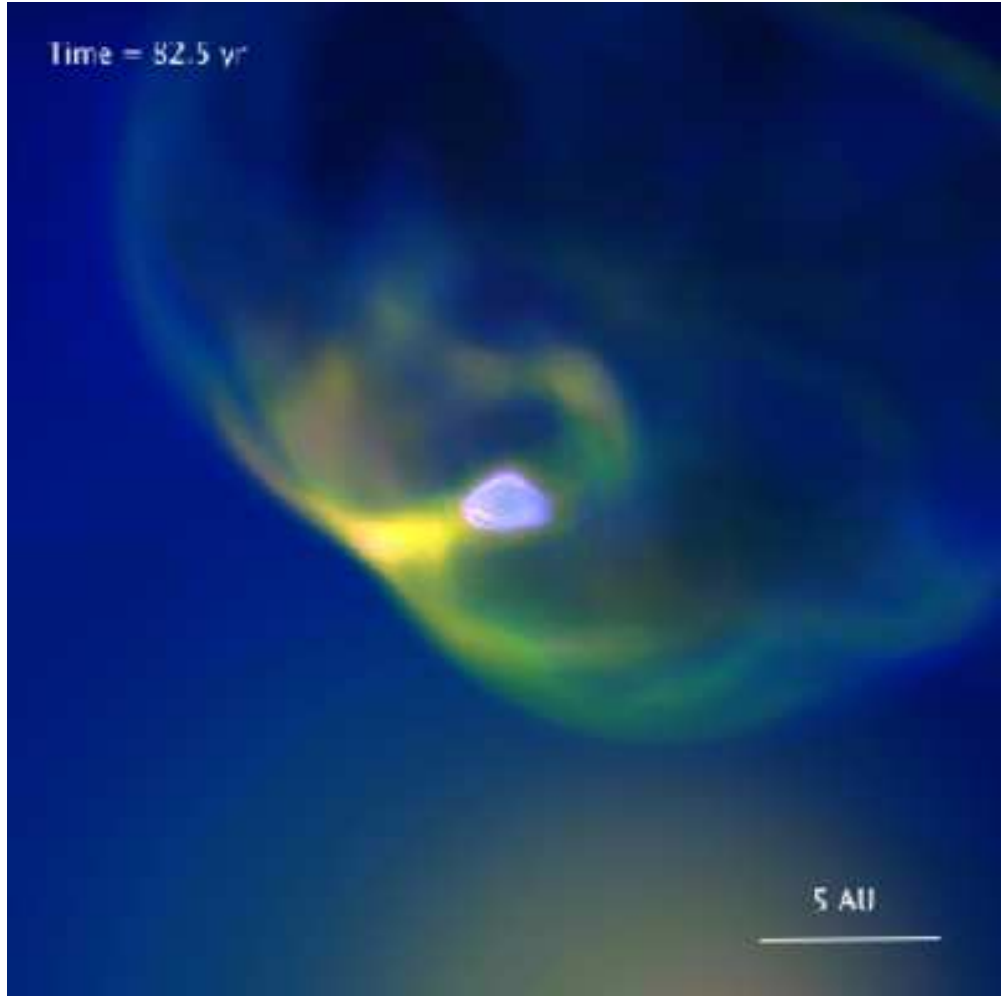}
\end{center}
\caption{\label{fig_disc} 3D volume rendering of gas density around the WD for model~M6. 
The white region represents the disc-like accretion flow around the WD 
and the yellow region on the left indicates the colliding shock front from the RG wind. 
The RG cannot be seen in this figure, but the bright region at the bottom 
shows the location of the RG.}
\end{figure}

The WD gravitationally deflects matter in its vicinity.  
Figure~\ref{fig_accretions} shows the mass within control surfaces for different radii 
for our models M6 and M7. 
It is clear that the enclosed mass increases for the first few years 
and then reaches stability after about 10~yr. 
Since the RG wind speed is higher than the orbital speed of the WD, 
the WD experiences a strong wind which limits the amount of mass enclosed within the 
control surface. 
Therefore, we observe no significant wind-capture disk around the WD as reported by \cite{2008A&A...484L...9W}, 
since the wind speed relative to the orbital speed is much higher than that in RS~Oph. 
However, we do observe a small, elongated, disk-shaped structure around the WD with a size about 5~AU 
(See Figure~\ref{fig_disc}).   
This result is consistent with the fast simulation case ($v_{\rm wind} = 60$~km~s$^{-1}$) of RS~Oph 
in \cite{2010ASPC..429..173W}, since the orbital speed of RS~Oph ($v_{\rm orb} \sim 15$~km~s$^{-1}$) 
is much faster than that in V407~Cyg ($v_{\rm orb} \sim 5$~km~s$^{-1}$), 
giving a similar orbital to wind speed ratio of $v_{\rm orb}/v_{\rm wind} \sim 1/4$. 
We note, however, that the enclosed mass is not gravitationally bound to the WD. As 
the WD ($R_{WD} \sim 4 \times 10^8$~cm) is not resolved in our simulations, we cannot  
determine the mass accreted by the WD.

For $r < 3 -5$~AU, the enclosed mass is enhanced up to $\sim 3-4$ times, 
independent of the radius of the control surface and the mass loss rate. 
Qualitatively, the models M6 and M7 behave similarly, 
but the density is one order of magnitude higher in model~M6 than in model~M7.
In addition, a short-period oscillation ($P \sim 2$~yr) in the enclosed mass is also observed. 
This prompt oscillation reflects the dynamical timescale around the WD 
due to the interaction between the colliding RG wind  and the turbulent flow behind the WD.

\begin{figure}
\begin{center}
\epsscale{1.0}
\plotone{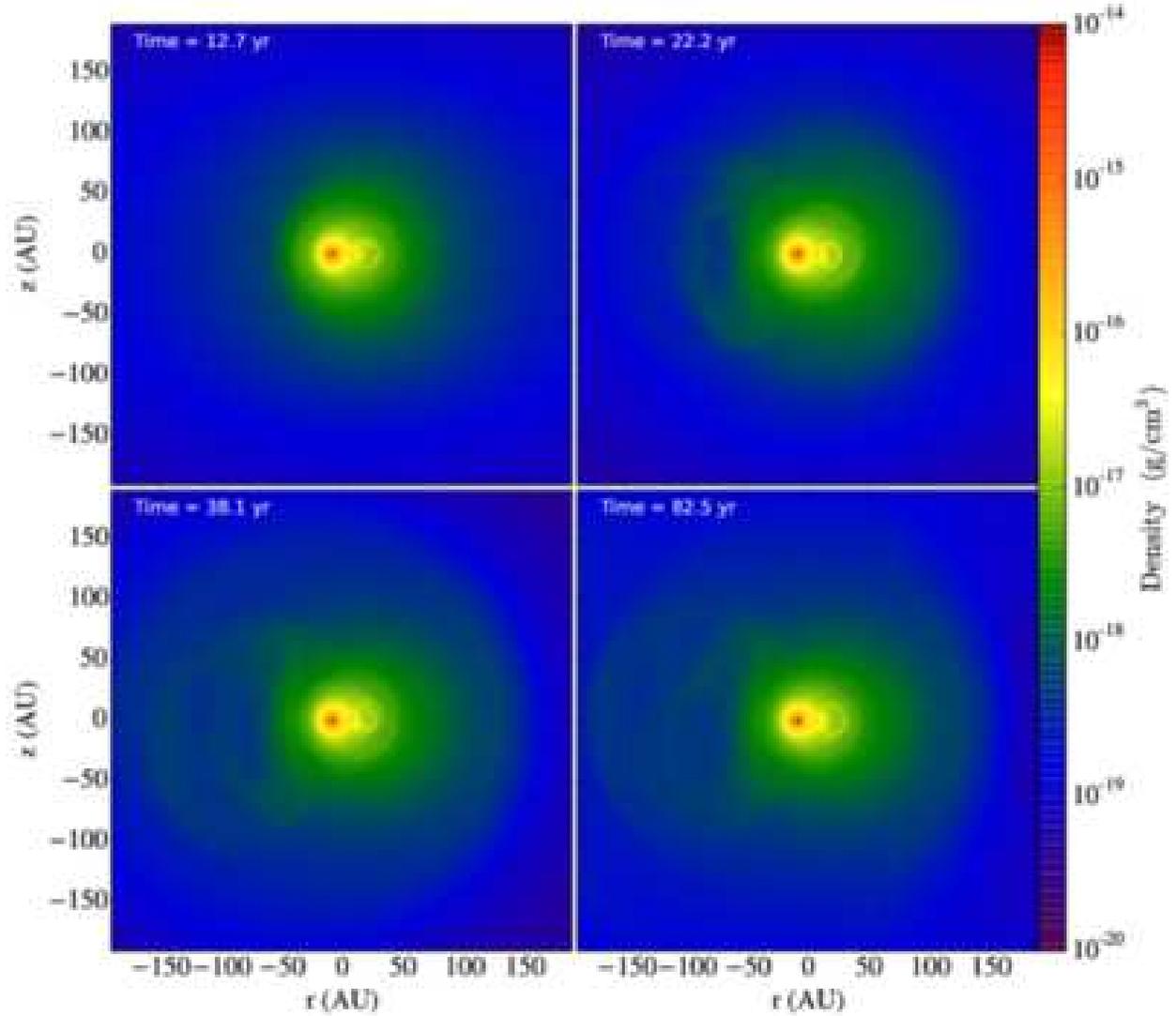}
\end{center}
\caption{\label{fig_M6_dens_r-z} 
Gas density projection in the $r-z$ plane perpendicular to the orbital plane
for model~M6 at different labeled times, 
where the $r$-axis represents the axis passing through the RG and the WD.
Note that the origin of the $r-$axis in this plot is at the center of mass, 
which is different from the simulation coordinate origin.   
In this figure, the RG is located on the left.
The color scale indicates the logarithm of the gas density in g~cm$^{-3}$.}
\end{figure}

\begin{figure}
\begin{center}
\epsscale{1.0}
\plotone{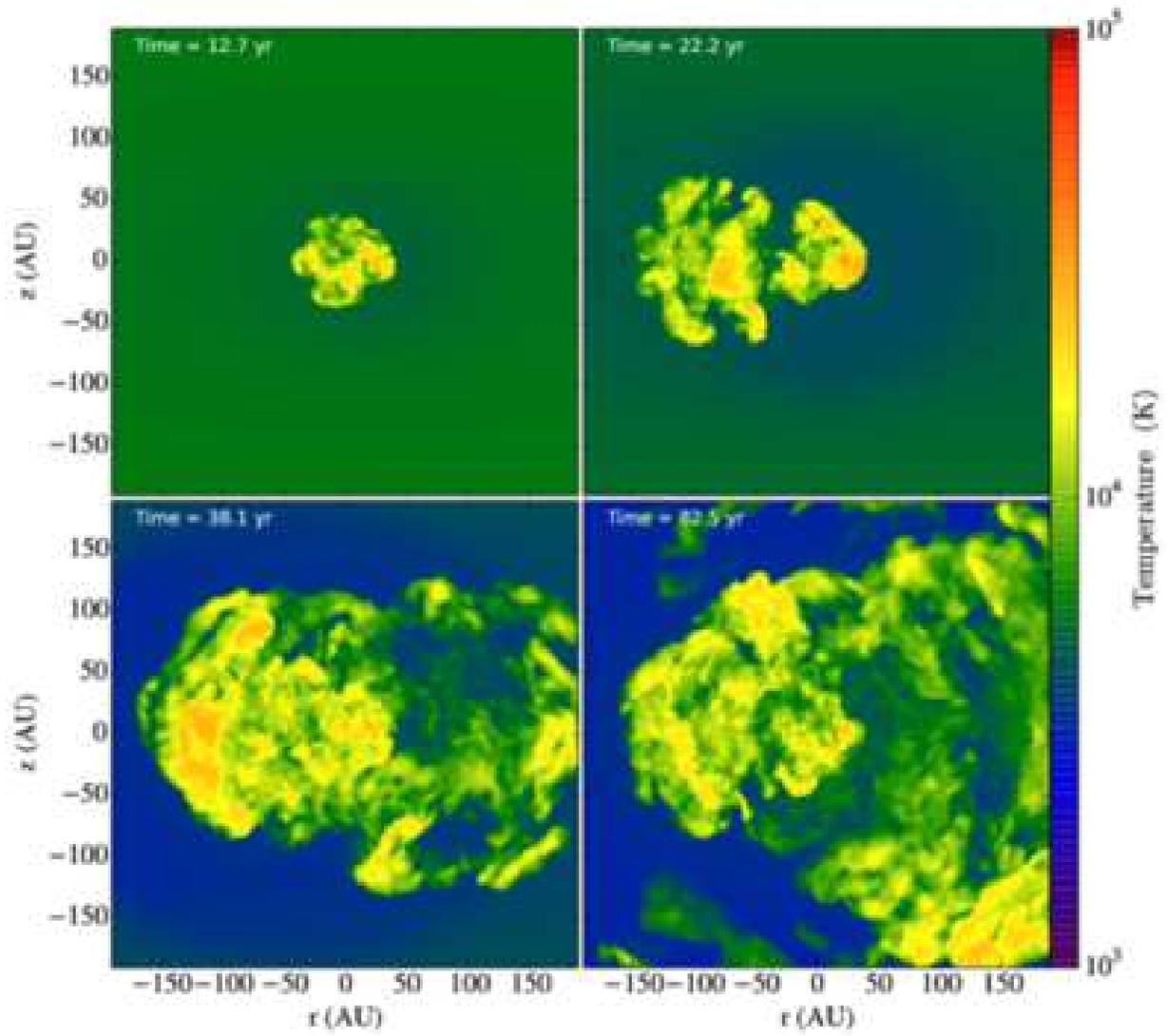}
\end{center}
\caption{\label{fig_M6_temp_r-z} Similar to Figure~\ref{fig_M6_dens_r-z} but for gas temperature.}
\end{figure}

The CDE, including the spiral arm, is concentrated in the equatorial plane. However, 
the CDE does not form a thin disk, but instead shows a mild anisotropy.
This is due to a relatively large pressure scale height ($H_{\rm p} \sim c_s/\Omega$) in V407~Cyg,
where $\Omega$ is the angular velocity.
Figures~\ref{fig_M6_dens_r-z} and \ref{fig_M6_temp_r-z} show the gas column density and gas 
density-weighted temperature distribution as viewed from within the orbital plane.  The density enhancement 
is only about  $\sim 20-40 \%$, but the temperature can be heated to $\sim 20,000$~K. 
The opening angle of the spiral arm is about $\sim 30 -50$ degrees. 
As described above, when the RG wind collides with the WD, some gas periodically puffs out from the WD 
and forms the ring-like density distribution in Figure~\ref{fig_M6_dens_r-z}.

\begin{figure}
\begin{center}
\epsscale{0.8}
\plotone{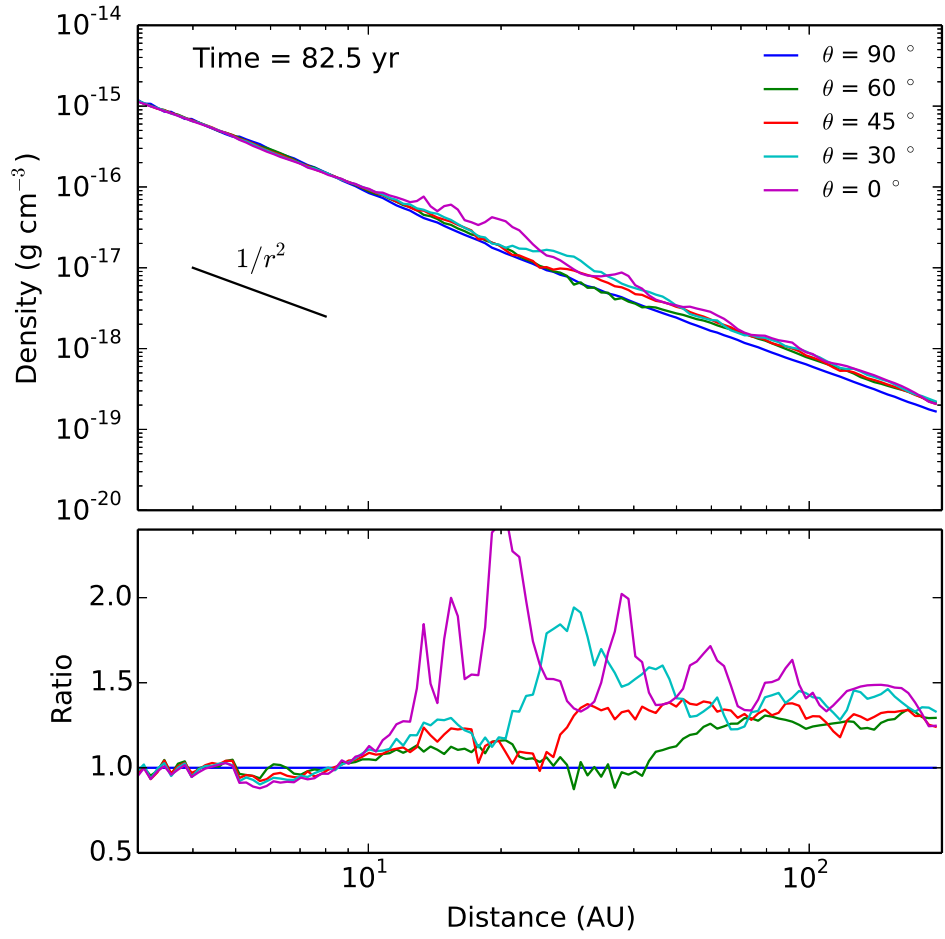}
\end{center}
\caption{\label{fig_M6} 
Top: Averaged density profile of model~M6 with different inclination angles as a function of radius at $82.5$~yr. 
Bottom: The ratio of the density at different inclination angles to that along the rotation axis.}
\end{figure}

\begin{figure}
\begin{center}
\epsscale{0.8}
\plotone{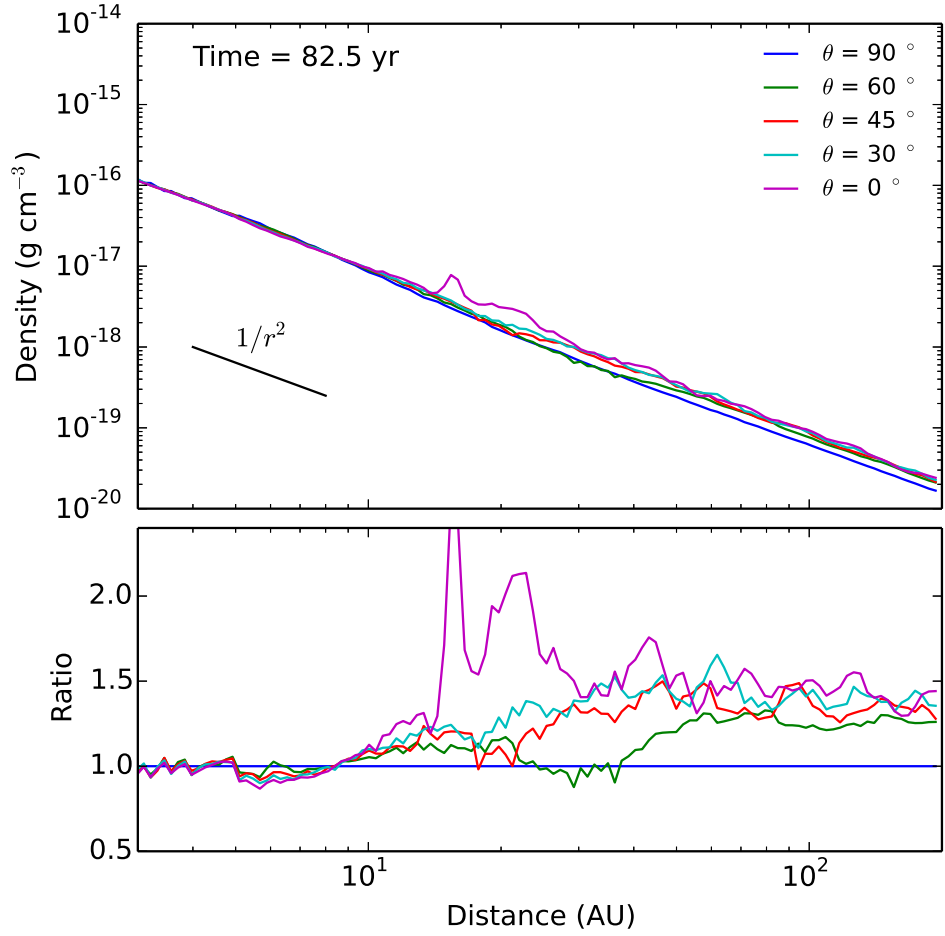}
\end{center}
\caption{\label{fig_M7} Similar to Figure~\ref{fig_M6} but for model~M7.}
\end{figure}

Figure~\ref{fig_M6} and \ref{fig_M7} show the angle-averaged density profiles of models~M6 and M7  
within surface of cones having different inclination angles.  
Each density profile is averaged from 180 uniformly distributed radial rays in a given inclination angle 
and each radial ray is sampled from the surface of the RG to the boundary of the simulation box.
From the upper panels, the averaged density profiles still follow the $1/r^2$ distribution,
but with about $20-40\%$ variation for different inclination angles. 
It is also clear that the density distribution at about $ r < 70-80$~AU
is highly concentrated in the region with $\theta < 60^\circ$. 
Beyond this scale, the asymmetry is small and only affected at high inclination angles. 

We have also performed a low-resolution run by reducing the maximum AMR level to 8 and
enlarging the finest zone size to $\Delta x_{\rm min} = 5.86 \times 10^{12}$~cm, 
corresponding to a factor of 2 larger than the standard resolution.
In this low-resolution run, the spiral arm qualitatively looks the same as in the standard run, 
and the averaged density distribution looks the same as well, 
but the turbulence is slightly suppressed due to a higher numerical viscosity.
The enclosed mass within the same control surface ($r< 5 R_c$) is about $20\%$ higher than the standard run.

It should be noted that in our model we assume a constant RG wind speed during the whole quiescent-phase simulation. 
However, in reality the RG wind is driven by radiation pressure on dust formed above the stellar surface during 
the Mira giant's pulsation \citep{2000ARA&A..38..573W}.  The wind acceleration zone could extend to about 
$5-10 R_{\rm RG}$ \citep{1988ApJ...329..299B}, which is comparable to the binary separation in V407~Cyg.
Therefore, the wind speed could be less than what we have assumed when it collides with the WD, 
implying that our estimate of the CDE could be low.

\subsection{Eruption Phase \label{sec_eruption}}

After running for two orbital periods in the quiescent-phase simulations, we impose a nova eruption 
on the WD.  We perform two eruption simulations with two different eruption energies ($E=1.2 \times 10^{43}$~erg 
and $E=1. 2 \times 10^{44}$~erg) and ejecta masses ($M_{\rm ej} = 10^{-7} M_\odot$ and $M_{\rm ej} = 10^{-6} 
M_\odot$) for each quiescent-phase simulation.  
Since we assume that the averaged ejecta speed is $\langle v_{\rm ej} \rangle= 3,000$~km/sec
and the kinetic energy ($E_{\rm kin} = \frac{1}{2} M_{\rm ej} \langle v_{\rm ej} \rangle^2$) 
is three-fourths of the total eruption energy at the onset of eruption \citep{1996ApJ...471..279D}, 
a higher eruption energy implies a larger ejecta mass in that eruption, and vice versa.  
In total, we have four different sets of 3D eruption simulations as described in 
Table~\ref{tab_simulations}. In Section~\ref{sec_description}, we adopt model M6E44 as the reference simulation 
to describe the overall evolution in the eruption phase.  We also discuss the shock evolution in all our models 
in Section~\ref{sec_shock}.  

\subsubsection{A Qualitative Description of the Evolution after the Nova Eruption \label{sec_description}}

\begin{figure}
\begin{center}
\epsscale{1.0}
\plotone{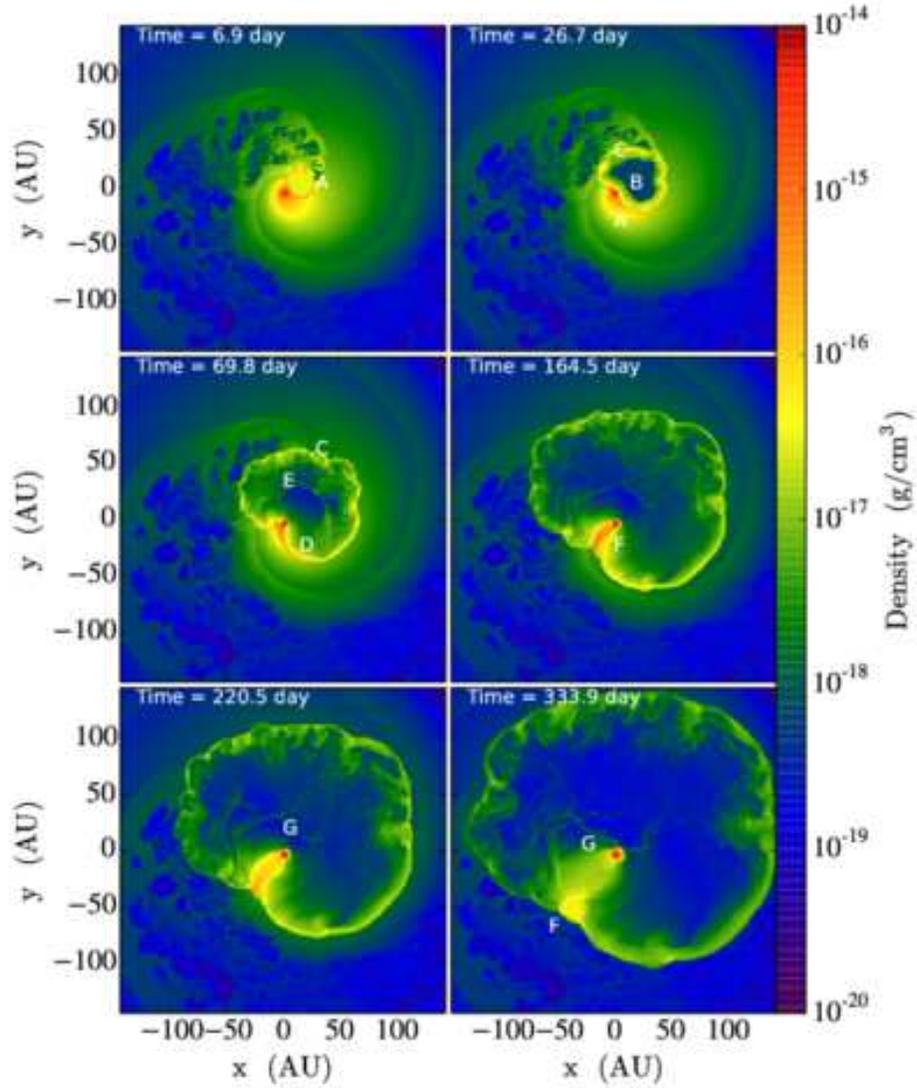}
\end{center}
\caption{\label{fig_M6E44_dens_x-y} 
Similar to Figure~\ref{fig_M6_dens_x-y} but for the eruption phase of model~M6E44. 
The labeled times indicate the time after a nova eruption. 
Labels are important features that are described in Section~\ref{sec_eruption}.}
\end{figure}

\begin{figure}
\begin{center}
\epsscale{1.0}
\plotone{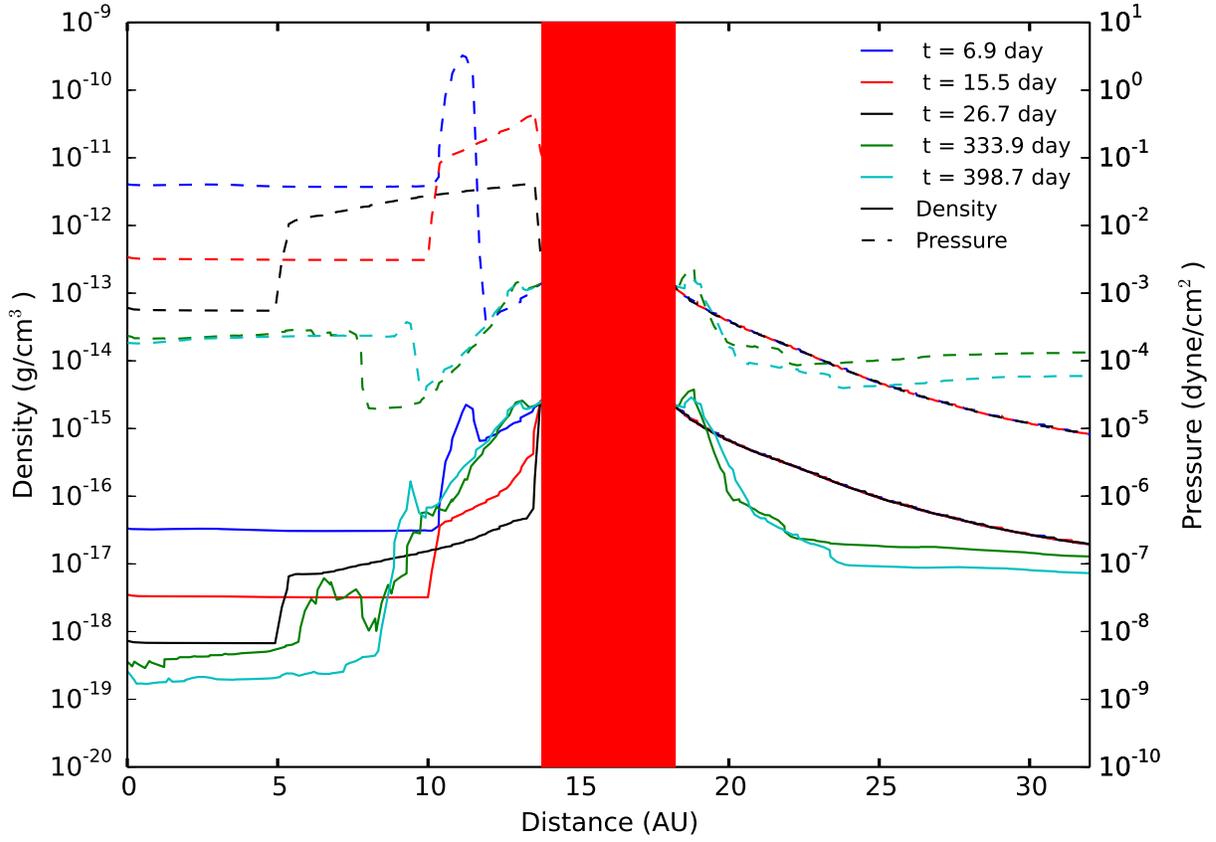}
\end{center}
\caption{\label{fig_ray_M6E44} Density (solid lines) and pressure (dashed lines) profiles along the line extending from the WD to the RG for model ~M6E44. 
Different colors represent different times after the RN eruption. The red shaded region shows the location of the RG.}
\end{figure}

Figure~\ref{fig_M6E44_dens_x-y} shows a typical gas density distribution for V407~Cyg in the orbital plane after a 
nova eruption.  In this case (model~M6E44 in Table~\ref{tab_simulations}), the mass loss rate of the RG wind is 
$M_{\rm wind} = 10^{-6} M_\odot$~yr$^{-1}$, the ejecta mass is $M_{\rm ej} = 10^{-6} M_\odot$, and the eruption 
energy is $E=1. 2 \times 10^{44}$~erg.  At about one week after the eruption, the forward shock sweeps out the 
RG wind and the spiral arm (label~A), and forms a reverse shock propagating backward to the WD (label~B). 
The two shocks are separated by a contact discontinuity.  At $\sim 15$~days, the forward shock reaches the 
RG.  The propagation of the forward shock and the reverse shock can be seen in Figure~\ref{fig_ray_M6E44}.
Note that although we assume the RG completely absorbs the forward shock when it reaches the surface of the 
RG, it does not affect the propagation of the reverse shock from the eruption. 

Subsequently, Rayleigh-Taylor instabilities quickly develop (label~C in Figure~\ref{fig_M6E44_dens_x-y}). 
After the forward shock passing through the RG, the RG wind and the CDE are shocked and heated (label~D).
These hot and dense plasma could be the source of X-ray emission \citep{2012MNRAS.419.2329O}.
The self-interaction of Rayleigh-Taylor bubbles and the interaction of the forward shock with the spiral 
arm produce additional reverse shocks as well.  When these reverse shocks collide with each other, 
filament-like shocks form at about 70~days after the eruption (label~E). 
 
\begin{figure}
\begin{center}
\epsscale{1.0}
\plotone{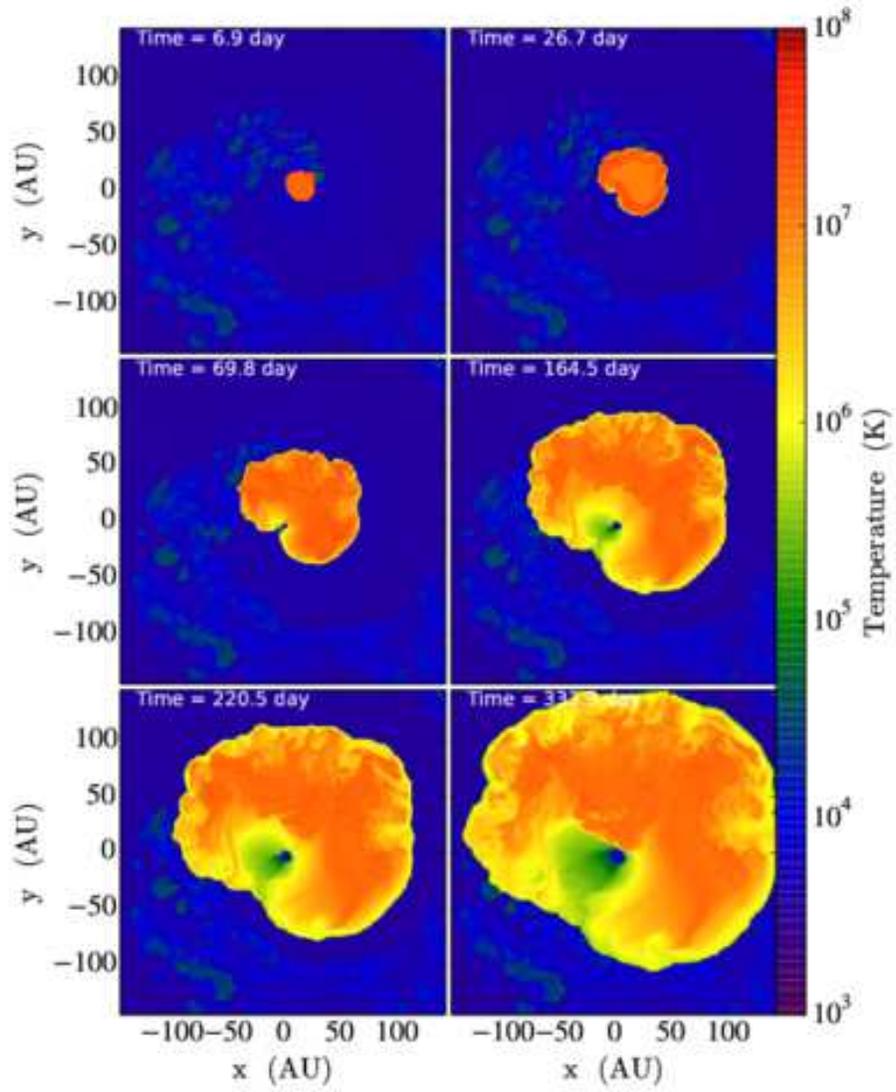}
\end{center}
\caption{\label{fig_M6E44_temp_x-y} 
Similar to Figure~\ref{fig_M6_temp_x-y} but for the eruption phase of model~M6E44. }
\end{figure}

After 70~days, the forward shock has slowed down because of the positive density gradient toward the RG, 
forming a heart-shaped structure.  Once the forward shock has passed 
the RG, a high-density tail at the back side of the RG is formed (label~F).  
Within the forward shock, several filaments are 
formed and destroyed repeatedly due to the interaction of the reverse shocks (label~G). The turbulent 
spiral arm is completely destroyed by the forward shock.  The corresponding gas temperature distribution is shown 
in Figure~\ref{fig_M6E44_temp_x-y}.  It is clear that the temperature distribution is highly asymmetric
due to the RG wind interaction and Rayleigh-Taylor instabilities. 
High Mach number shocks are associated with the forward shock, 
starting from $M \gtrsim 10$ in the first week and then decreasing with time. 
Once the Rayleigh-Taylor instabilities have developed, the  
Mach number decreases to $M \sim 5$ and is maintained at this level for several months.

\begin{figure}
\begin{center}
\epsscale{0.48}
\plotone{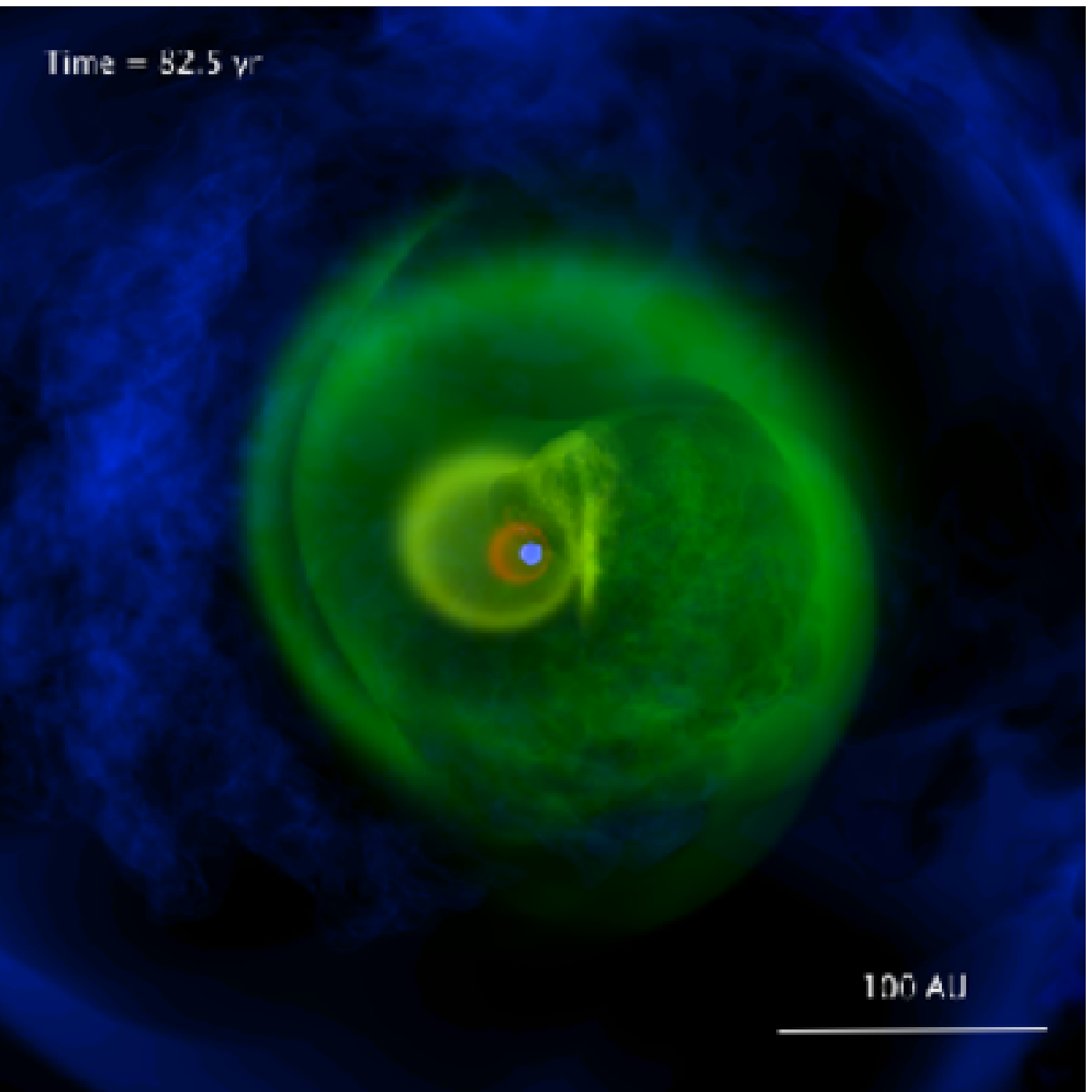}
\plotone{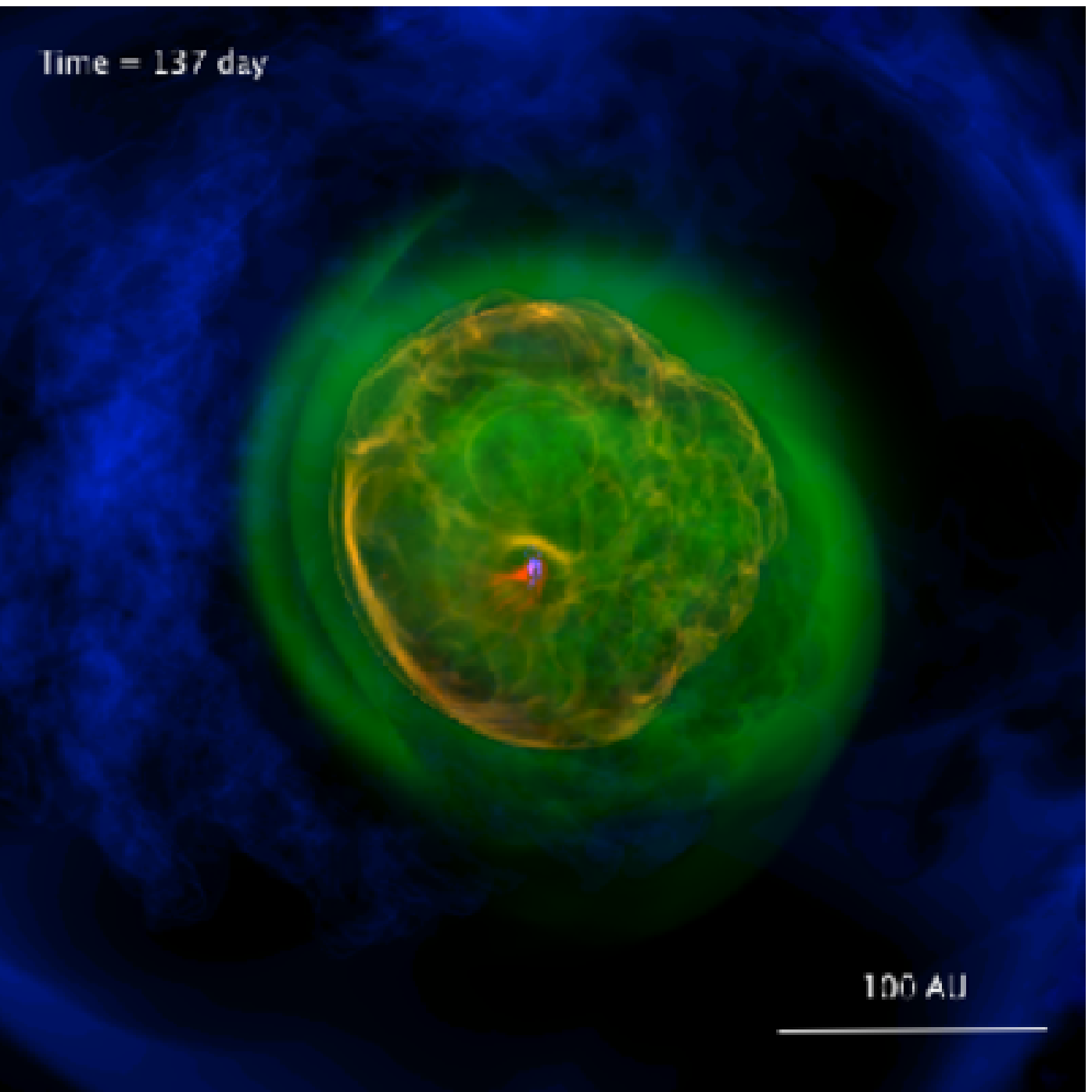}
\end{center}
\caption{\label{fig_vr_M6E44} 
3D volume rendering of gas density for models~M6 and M6E44 at different simulation phases. 
Left: In the quiescent phase (model~M6), the labeled time indicates the simulation time since the start of the simulation.
Right: In the eruption phase (model~M6E44), the labeled time indicates the simulation time after nova eruption.
Orange colors in the right panel show the location of the nova shock. 
Red and yellow colors represent high-density regions; green and deep blue colors denote low-density regions. 
(movies of simulations are available online).}
\end{figure}

A 3D volume rendering of the gas density distribution right before and after ($t=137$~day) a nova eruption is 
shown in Figure~\ref{fig_vr_M6E44}.  The orange color shows the location of the forward shock.  Several 
filaments and Rayleigh-Taylor bubbles can also be seen.  

\begin{figure}
\begin{center}
\epsscale{1.0}
\plotone{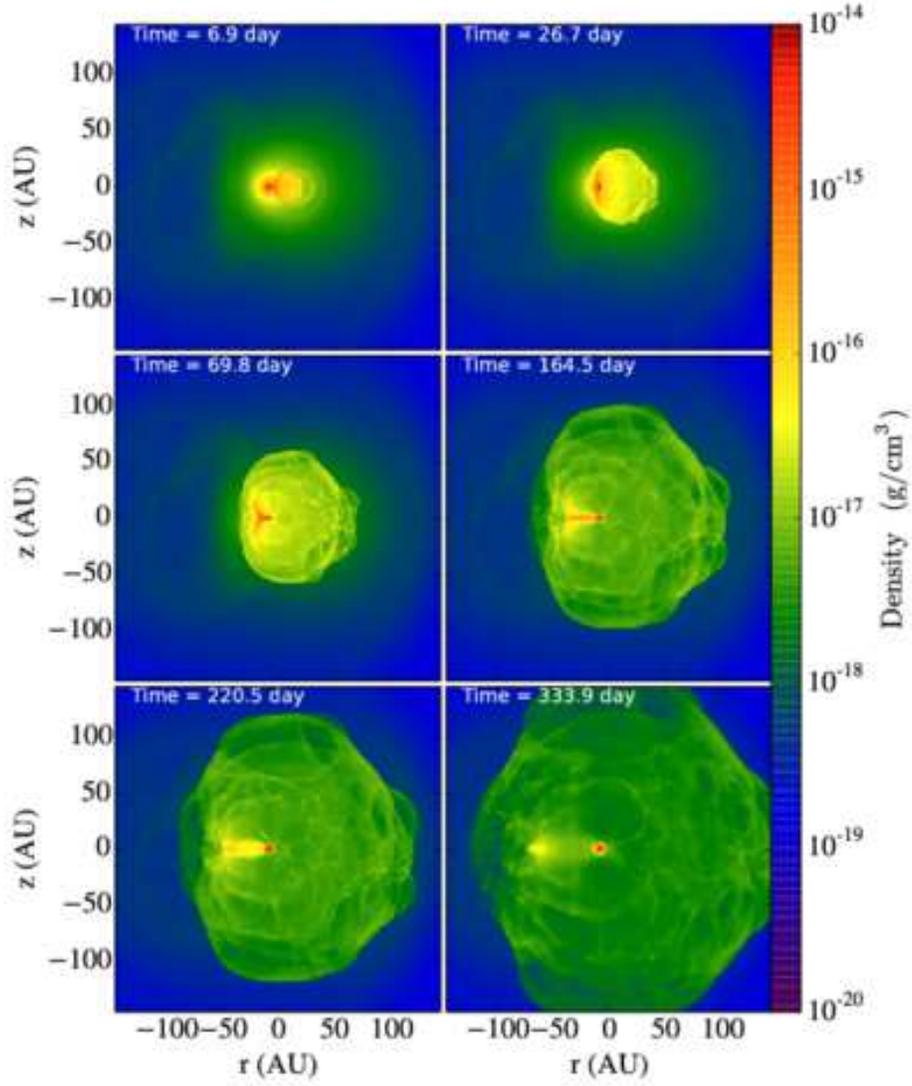}
\end{center}
\caption{\label{fig_M6E44_dens_r-z} 
Similar to Figure~\ref{fig_M6_dens_r-z} but for the eruption phase of model~M6E44. 
The labeled times indicate the time after a nova eruption.}
\end{figure}

\begin{figure}
\begin{center}
\epsscale{1.0}
\plotone{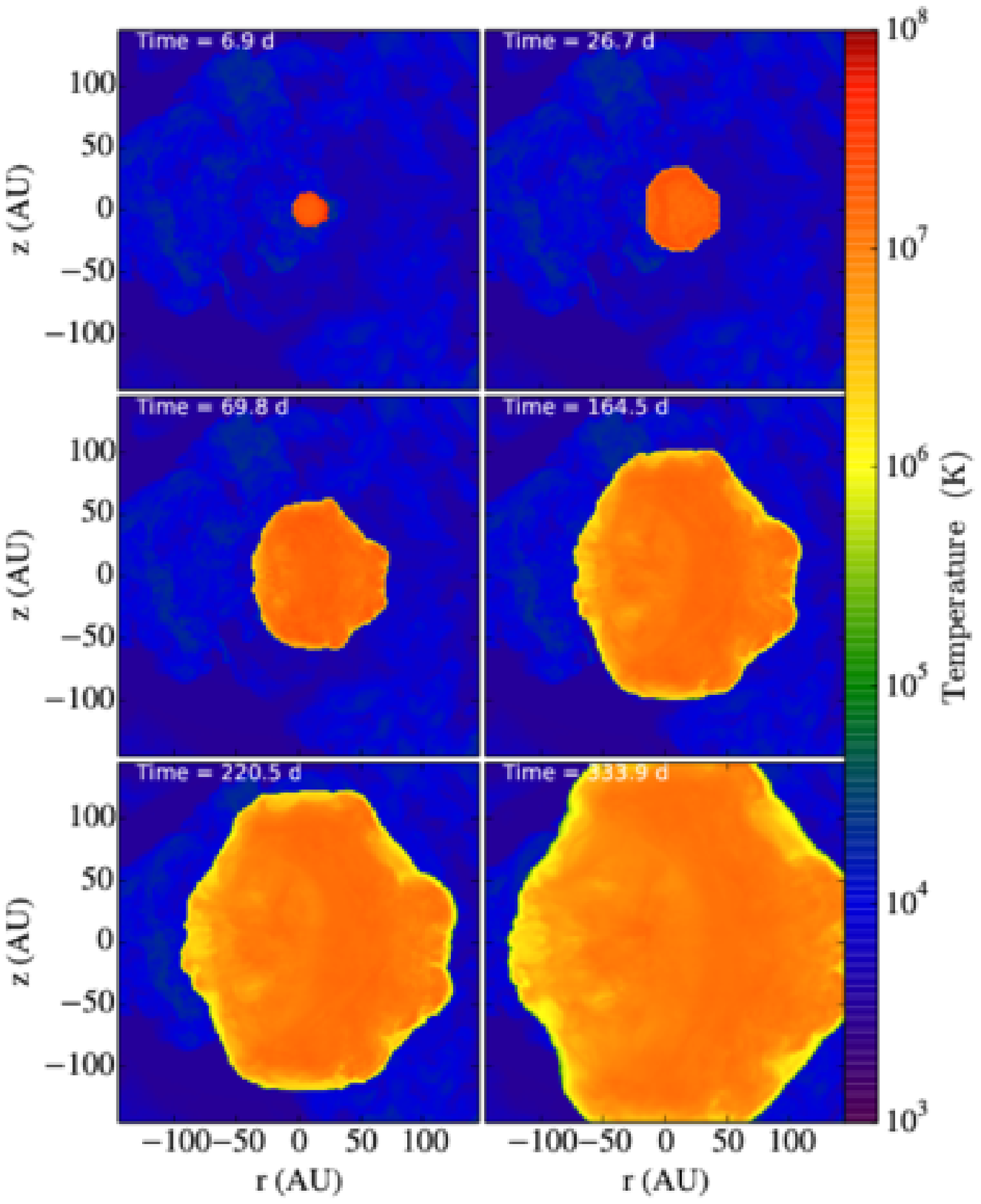}
\end{center}
\caption{\label{fig_M6E44_temp_r-z} Similar to Figure~\ref{fig_M6E44_dens_r-z} but gas temperature.}
\end{figure}

In Figure~\ref{fig_M6E44_dens_r-z} and \ref{fig_M6E44_temp_r-z}, we show the gas column density and gas 
density-weighted temperature distribution in the $r-z$ plane (edge-on view), to demonstrate the aspherical 
evolution of the shock radius due to the CDE.  As we described in the previous subsection, 
the CDE makes the density $\sim 20\%$ asymmetric at different inclination angles.  Therefore, 
although the ejecta speed is much higher than the orbital speed and the gravitational binding energy of the CDE 
is much less than the eruption energy, the ejecta speed greatly decreases in the equatorial direction 
due to the CDE and the RG wind.
The ratio of the shock radius in the poleward and equatorial directions is about $1.2 - 1.7$,
depending on the location of the RG. 

\cite{2012MNRAS.419.2329O} have performed simulations of the 2010 eruption of V407~Cyg with an 
artificial CDE. They found that the observed X-ray emission could be reproduced if there is a CDE 
around the WD, but is not required if the outburst energy and ejecta mass are near the upper 
end of the range for these characteristics for classical novae, which would be extreme for SyRN. 
Comparing the results of their best-fitting simulation model with CDE, E44.3-NW7-CDE6.3-L40 
with our results for model M7E44,
we find that their forward shock expands more spherically and about twice as fast as 
that found here. \cite{2012MNRAS.419.2329O} found that most X-ray emission originates from the shocked CSM.
The observed X-ray emission reaches a maximum at around 40 days after the eruption and declines after 50 
days \citep{2011A&A...527A..98S}.  
This corresponds to the time when the RG wind and the CDE are heated by the shock in 
our simulations (see label~D in Figure~\ref{fig_M6E44_dens_x-y}). 
A lower shock expansion velocity in our simulations 
in comparison with \cite{2012MNRAS.419.2329O} may affect the shape of simulated X-ray lightcurve.    
However, we note that the initial averaged ejecta velocity $\langle v_{\rm ej} \rangle \sim 9,000$~km~s$^{-1}$
in \cite{2012MNRAS.419.2329O} is probably too high 
based on recent observations \citep{2010Sci...329..817A}.

Since the nova eruption completely destroys the CDE and creates a cavity that contains the binary 
system at the end of the simulation, the subsequent circumstellar density distribution will need several years 
to decades to reestablish the original wind profile (Equation~\ref{eq_rho}).  The last recorded nova 
eruption of V407~Cyg prior to the 2010 eruption was in 1998.  Assuming that there were no undetected 
eruptions between the 1998 and 2010 eruptions and the RG wind has a constant wind speed of 20~km~s$^{-1}$,
the newly developed wind profile could only reach to about $\sim 50$~AU in distance. 
Therefore, our initial wind profile (Equation~\ref{eq_rho}) may not be valid at large distances, 
which may affect our shock evolution at late times ($t > 3$~months for Model M6E44).   
  
If we reduce the resolution by a factor of 2 as we did in the quiescent-phase simulations, we see no 
notable difference in the shock locations within $r<150$~AU, but the Rayleigh-Taylor instabilities 
are less obvious due to a higher numerical viscosity.  The filaments produced by the interaction of reverse 
shocks are slightly less clear than in the standard run, but overall the evolution does not change 
significantly. 
  
\subsubsection{Shock Evolutions \label{sec_shock}}

Understanding the shock evolution in nova eruptions is crucial, since the high-energy charged 
particles required for the nonthermal gamma-ray and radio emission are likely produced in
the shock front.  Overall, the simulation results for the shock positions and evolution during the eruption 
phase runs are qualitatively similar, but quantitatively very different.  Therefore, in this subsection 
we describe the shock evolution of our four eruption simulations in detail. 

\begin{figure}
\begin{center}
\epsscale{0.48}
\plotone{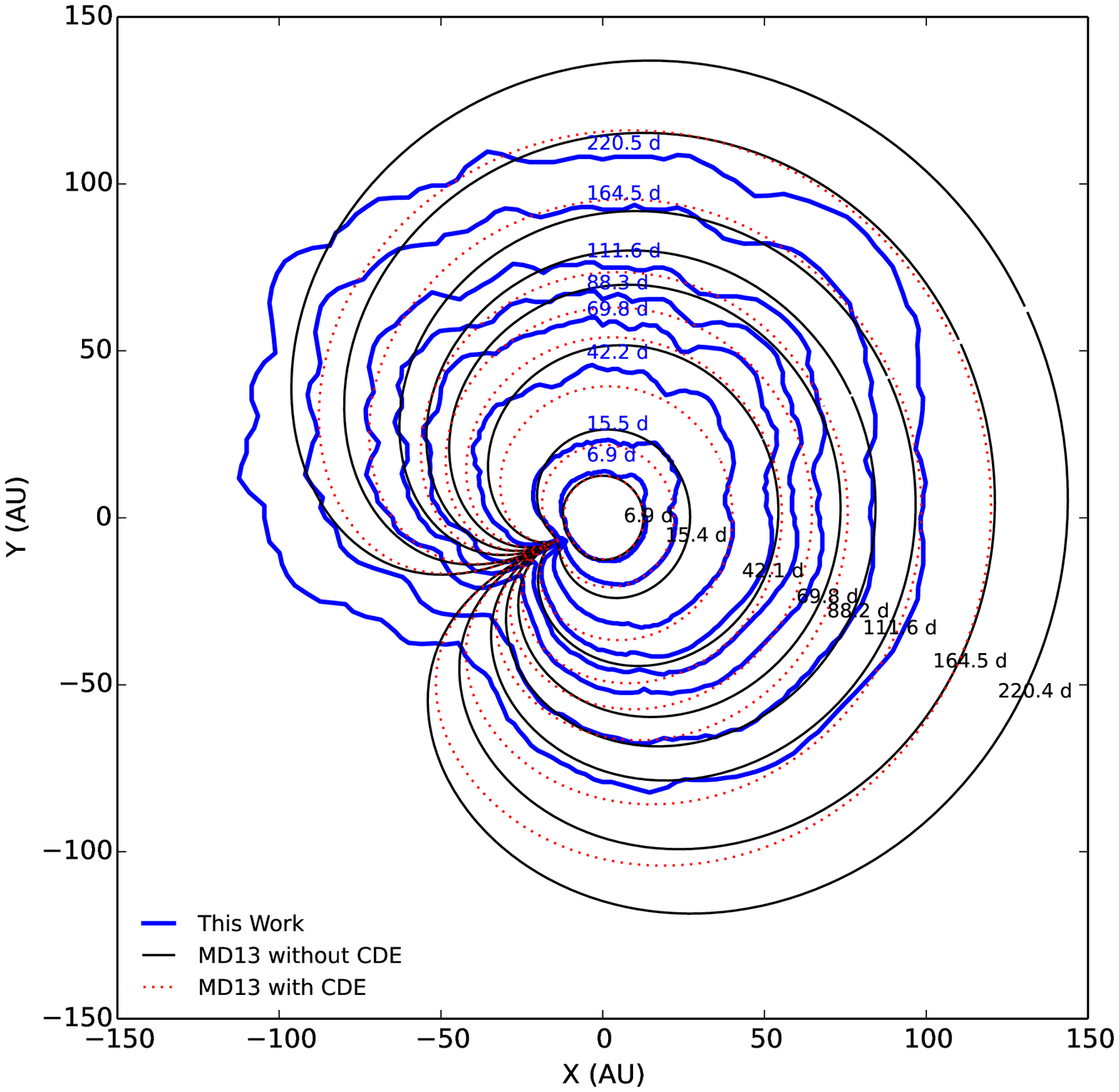}
\plotone{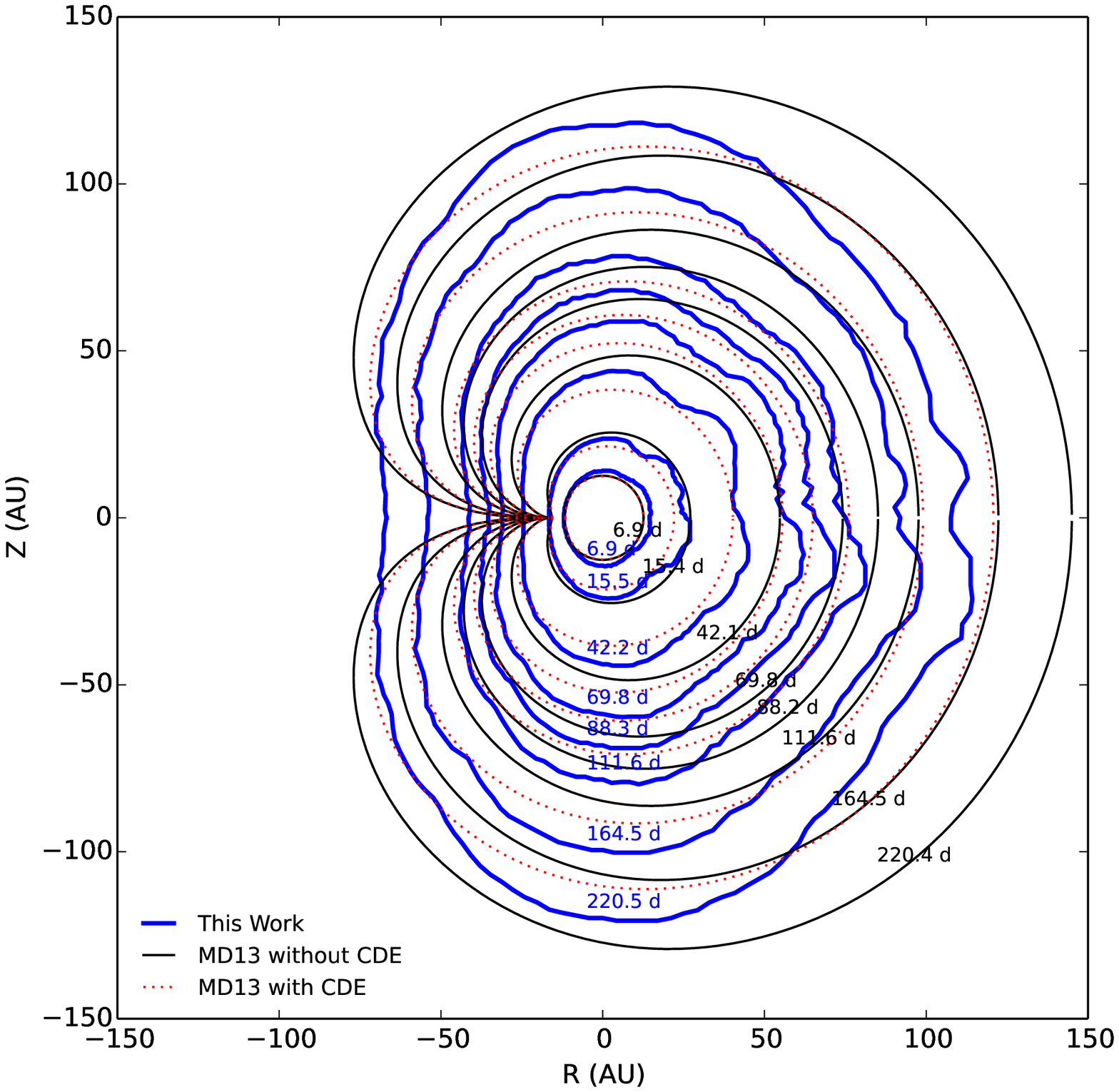}
\end{center}
\caption{\label{fig_shock_M6E44} 
Shock radius evolution. Left: in the orbital plane. 
Right: in the $r-z$ plane perpendicular to the orbital plane.
Blue contours indicate the shock radius evolution of model~M6E44 at different times.  
Black contours show a comparison of shock radius evolution with a simple analytical model (denoted 
by the label MD13) by \cite{2013A&A...551A..37M}.  Red dots represent the same simple analytical model, but 
with the CDE.}
\end{figure}

\begin{figure}
\begin{center}
\epsscale{0.48}
\plotone{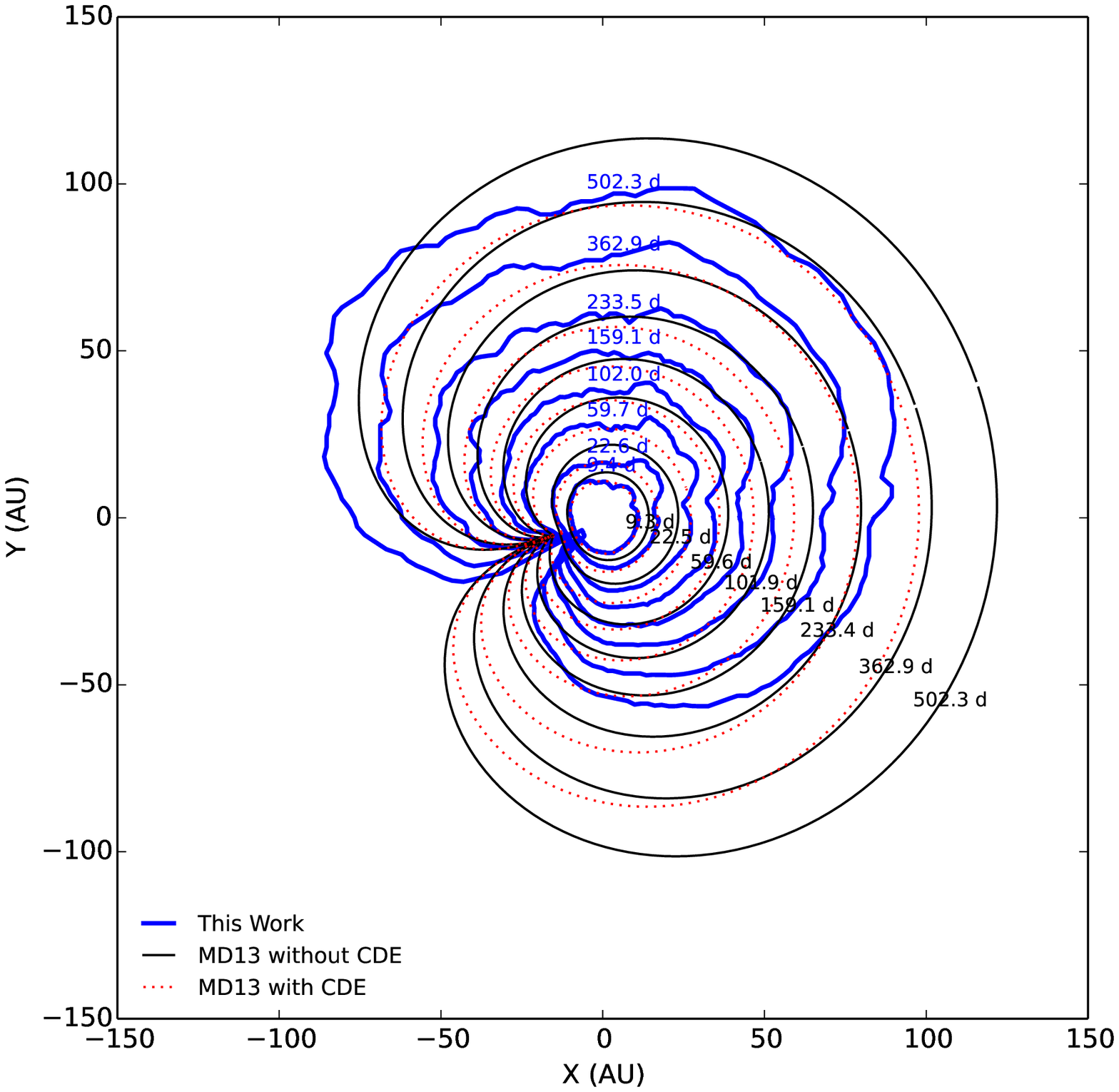}
\plotone{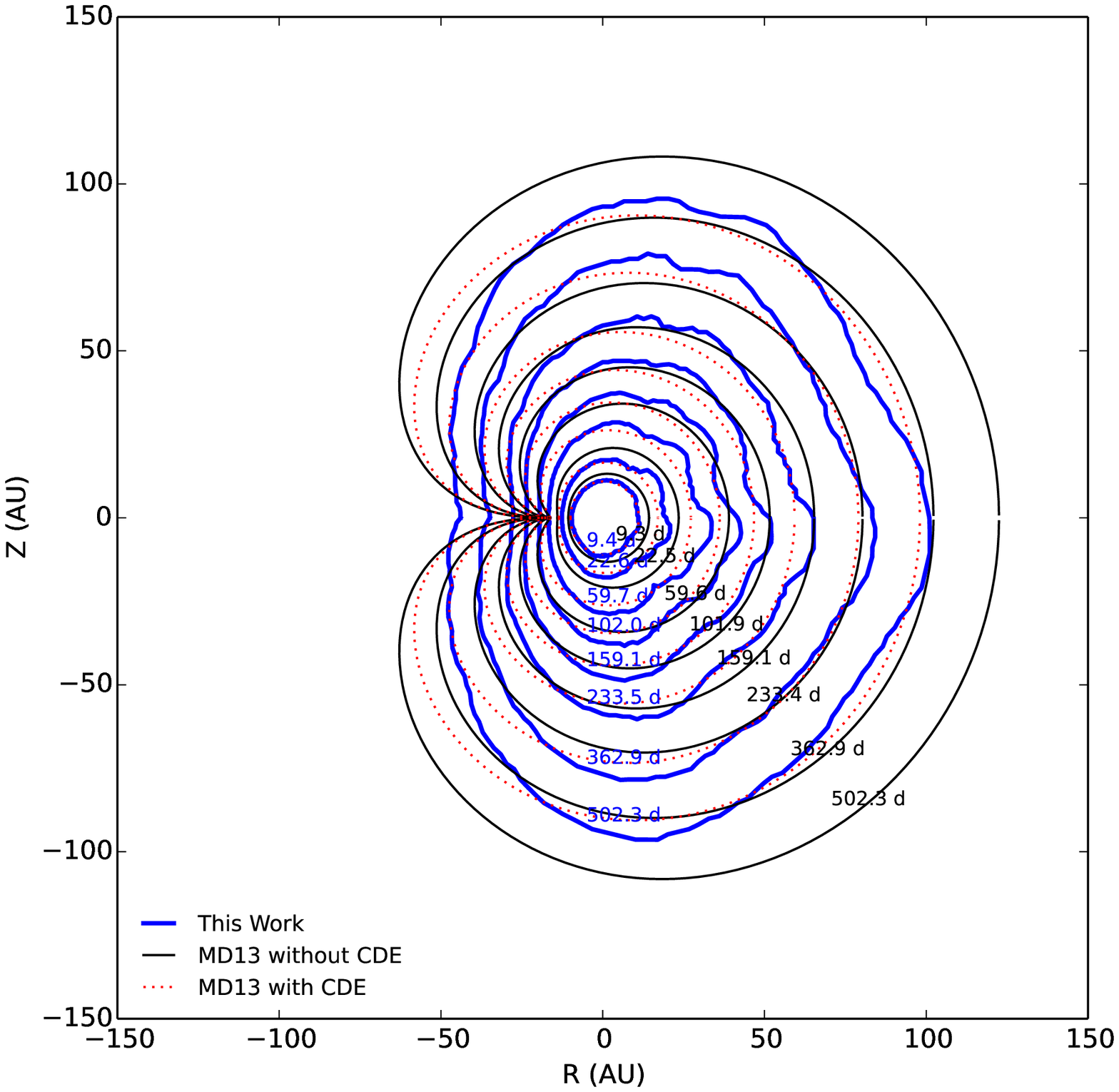}
\end{center}
\caption{\label{fig_shock_M6E43} Similar to Figure~\ref{fig_shock_M6E44} but for model~M6E43.}
\end{figure}

\begin{figure}
\begin{center}
\epsscale{0.48}
\plotone{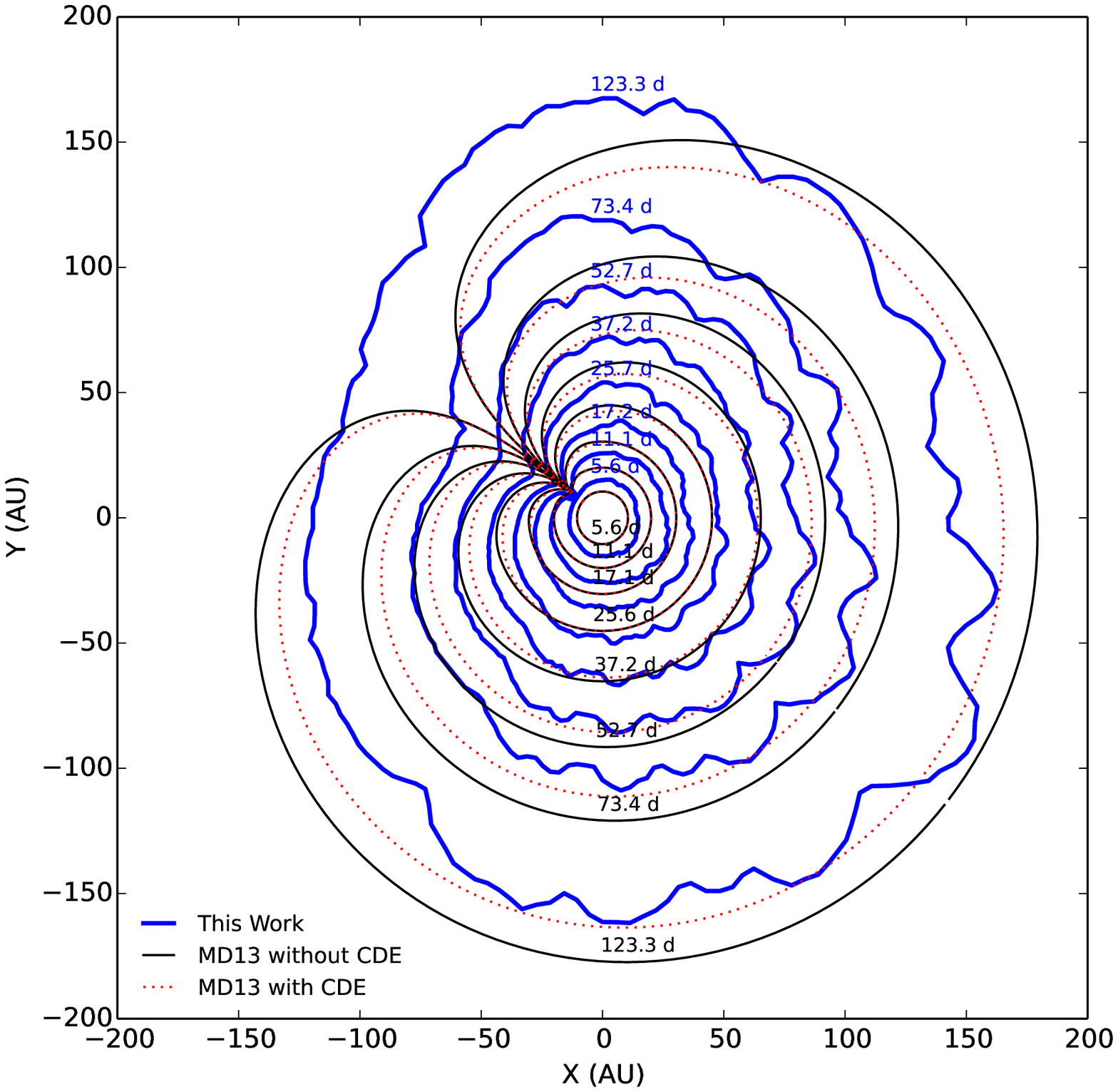}
\plotone{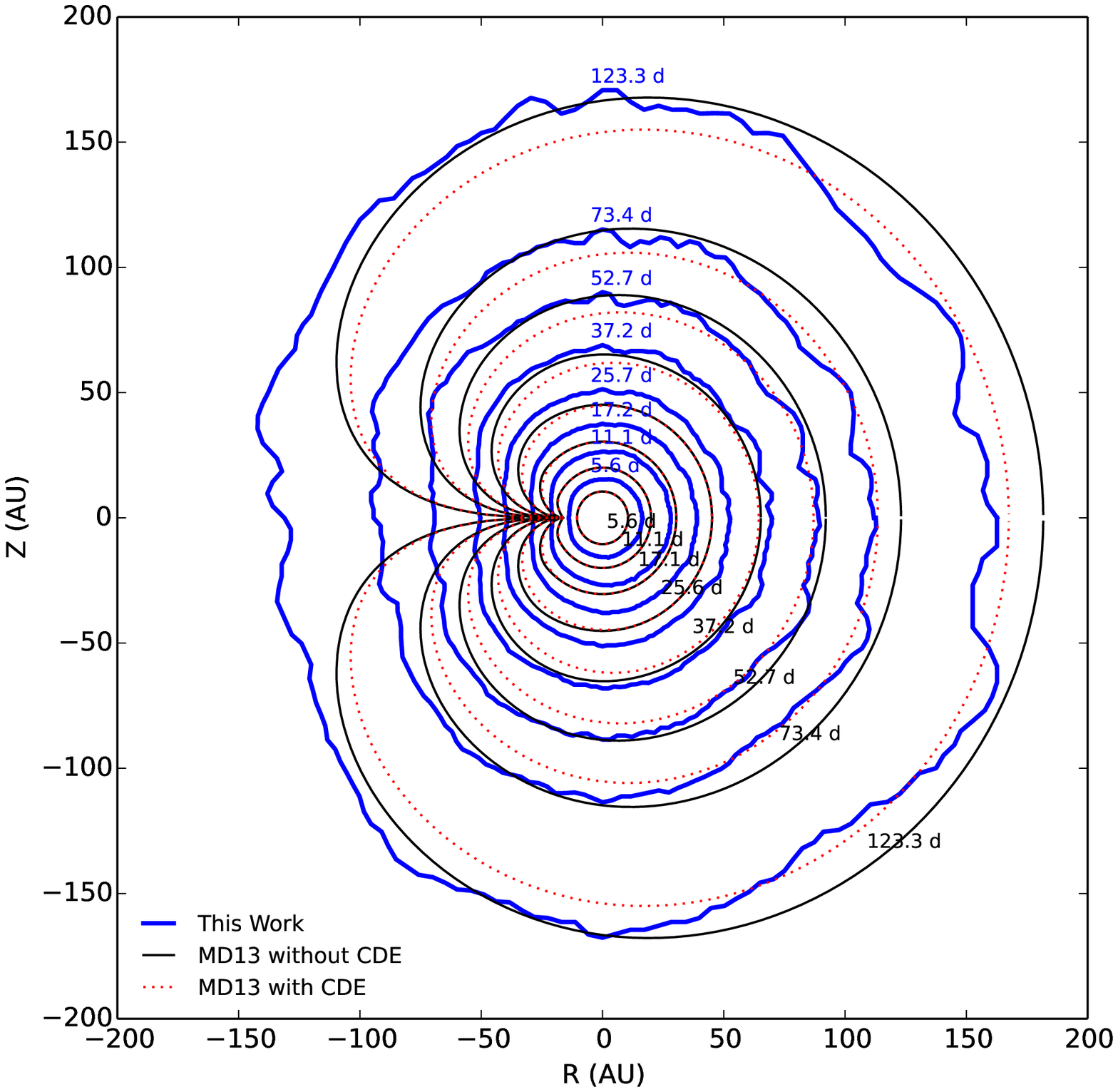}
\end{center}
\caption{\label{fig_shock_M7E44} Similar to Figure~\ref{fig_shock_M6E44} but for model~M7E44.}
\end{figure}

\begin{figure}
\begin{center}
\epsscale{0.48}
\plotone{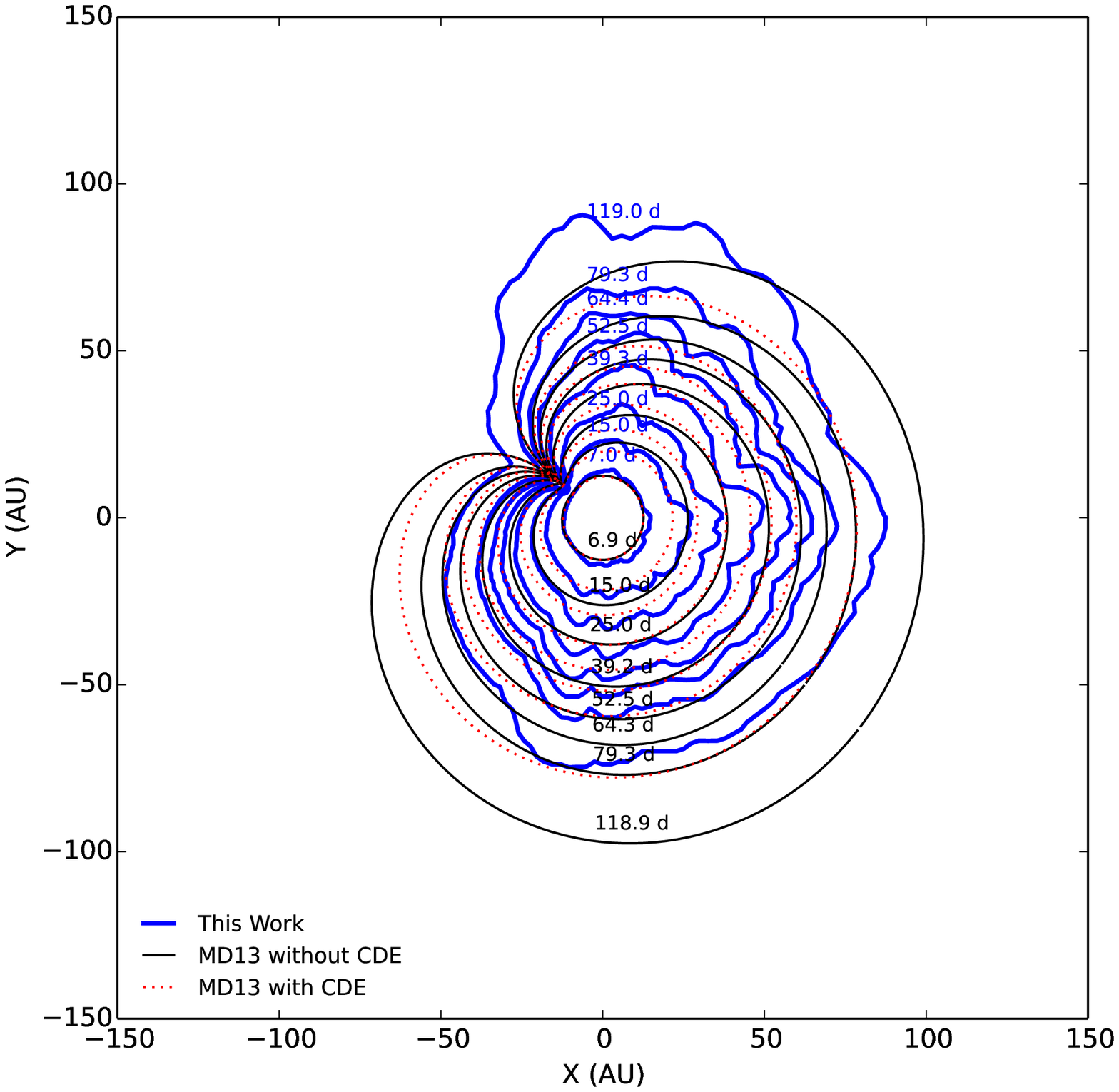}
\plotone{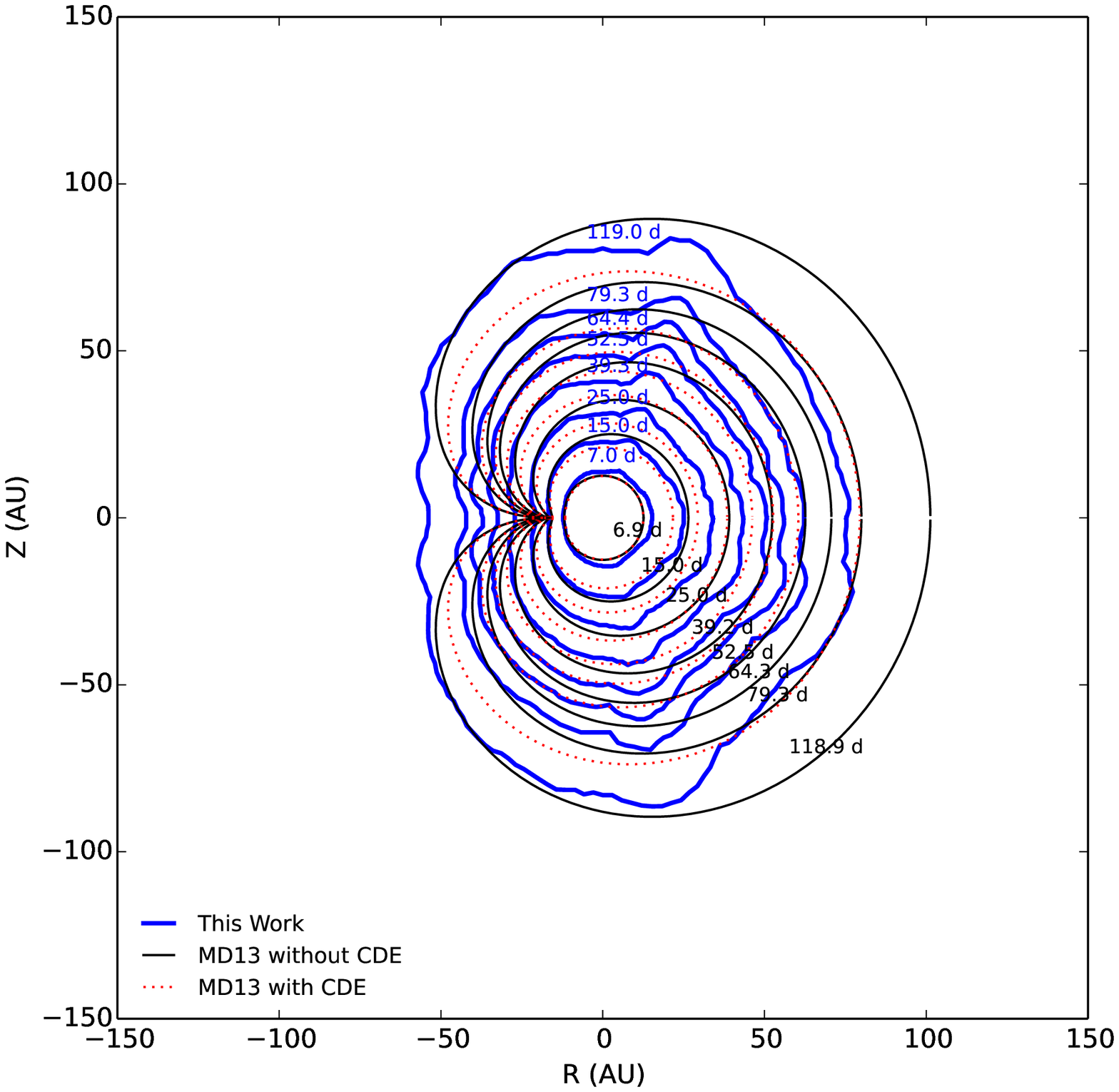}
\end{center}
\caption{\label{fig_shock_M7E43} Similar to Figure~\ref{fig_shock_M6E44} but for model~M7E43.}
\end{figure}

Figure~\ref{fig_shock_M6E44} to Figure~\ref{fig_shock_M7E43} show the shock evolution in all the eruption 
simulations.  At the very beginning of a nova eruption ($t < 1$~week), the eruption is nearly free-expanding 
and the shock location is nearly spherically symmetric.  Since the kinetic energy is the dominant 
form of energy at that time and we assume the initial averaged ejecta speed is the same in different runs, 
the shock radius evolution is similar in all cases.  However, later on, when the ejecta sweep up 
enough mass, the forward shock slows down due to momentum conservation.  The decline rate 
depends on the eruption energy and the RG mass loss rate.  For instance, if we lower the eruption energy 
or the ejecta mass by an order of magnitude (case~M6E43), the shock location would be slowed down by a 
factor of $\sim 2$ at $t \sim 200$~days (Figure~\ref{fig_shock_M6E43}).  Similarly, with the same eruption 
energy, the shock would propagate much faster if the surrounding CDE were less dense 
(Figure~\ref{fig_shock_M7E44}).  

\cite{2012ApJ...748...43N} and \cite{2013A&A...551A..37M} have described a simple semi-analytical model 
for the shock evolution by comparing the circumstellar mass distribution with the ejecta mass plus swept mass
using momentum conservation.  In their model, the density distribution is described by a 1D wind 
profile (Equation~\ref{eq_rho}) plus a CDE.  The CDE is assumed to be a gaussian distribution which can be 
described by 
\begin{equation}
\rho_{\rm CDE}(r,\theta) = \rho_{\rm 0, CDE} \exp \left(- \left(\frac{r \sin \theta}{b} \right)^2 - \left(\frac{r \cos \theta}{ l} \right)^2  \right),
\end{equation}
where $\rho_{\rm 0, CDE}$, $l$, and $b$ are density and length scale parameters.  We perform similar calculations 
by comparing the shock location of our 3D hydrodynamic simulations with the semi-analytical model described 
by \cite{2013A&A...551A..37M}. 

To find the most suitable sets of parameter values, we apply two constraints.
First, we assume the $\rho_{\rm 0, CDE}$ in model M6 is one order of magnitude higher than that in M7, 
i.e.~$\rho_{\rm 0, CDE}({\rm M6}) / \rho_{\rm 0, CDE}({\rm M7}) =10$, 
based on the analysis from Figure~\ref{fig_M6} and \ref{fig_M7}. 
Second, we assume the length-scale parameters $l$ and $b$ are equal, 
since we do not observe significant asymmetry in the local density enhancement around the WD.  
By applying these two constraints, a best-fit set of parameter values is:
for M7, $l=b=20$~AU and $\rho_{\rm 0, CDE} = 8 \times 10^{-18}$~g~cm$^{-3}$; and 
for M6, $l=b=20$~AU and $\rho_{\rm 0, CDE} = 8 \times 10^{-17}$~g~cm$^{-3}$.
The best-fit shock evolution is shown using the red dotted lines in Figures~\ref{fig_shock_M6E44} 
- \ref{fig_shock_M7E43}. 

The best fit model in \cite{2012MNRAS.419.2329O}, which can reproduce the X-ray lightcurve, corresponds
to their model E44.3-NW7-CDE6.3-L40.  It is characterized by an ejecta mass of $2 \times 10^{-7} M_\odot$, 
eruption energy $E = 2 \times 10^{44}$~erg, $\rho_{\rm 0,CDE} \sim 3 \times 10^{-18}$~g~cm$^{-3}$, and 
$b=l=40$~AU.  The CDE in this model is comparable to, in order of magnitude, with our model M7, but is less 
concentrated.  However, as described above, the ejecta speed in \cite{2012MNRAS.419.2329O} is much faster 
than adopted in our work.

\begin{figure}
\begin{center}
\epsscale{0.48}
\plotone{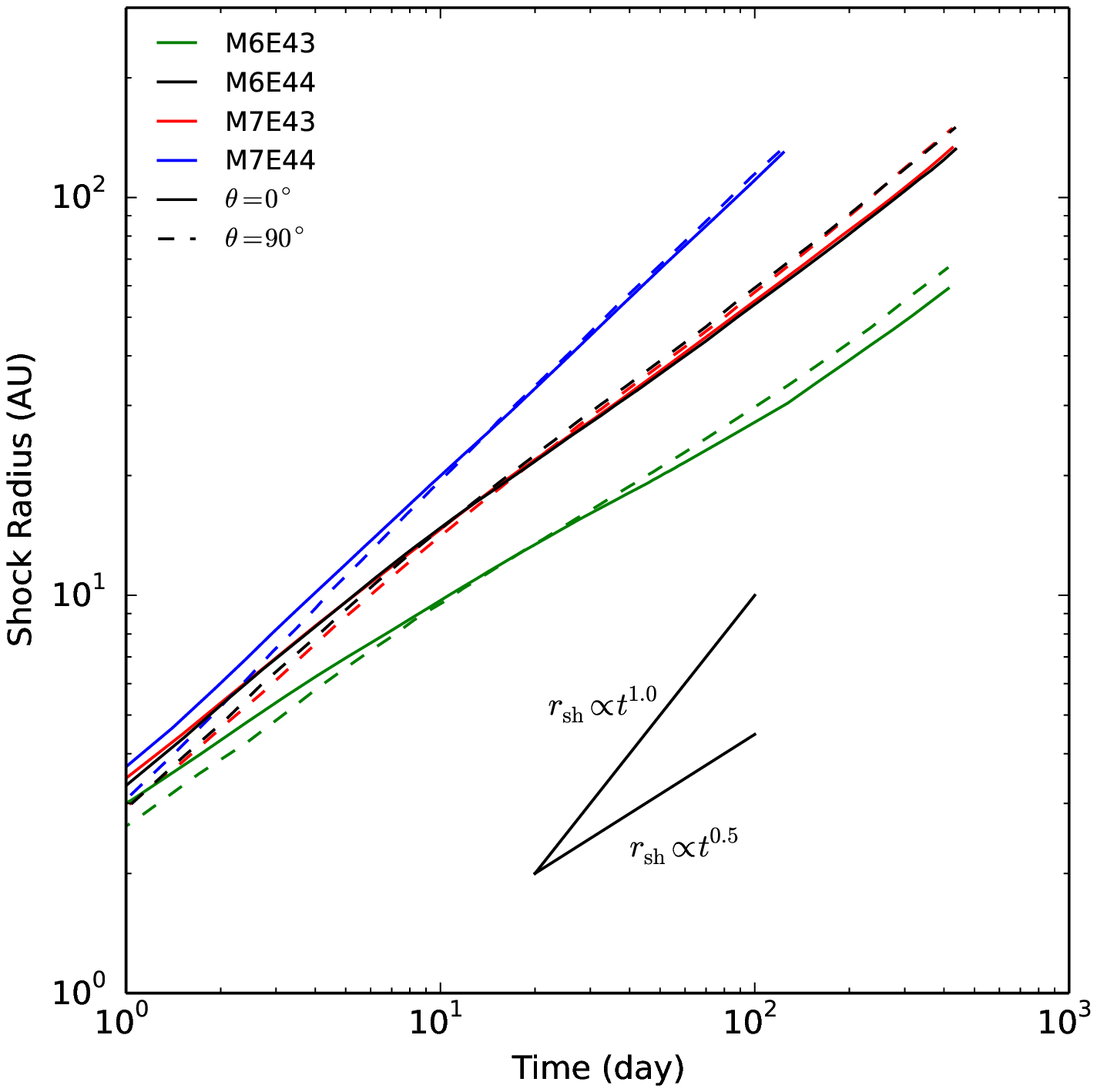}
\plotone{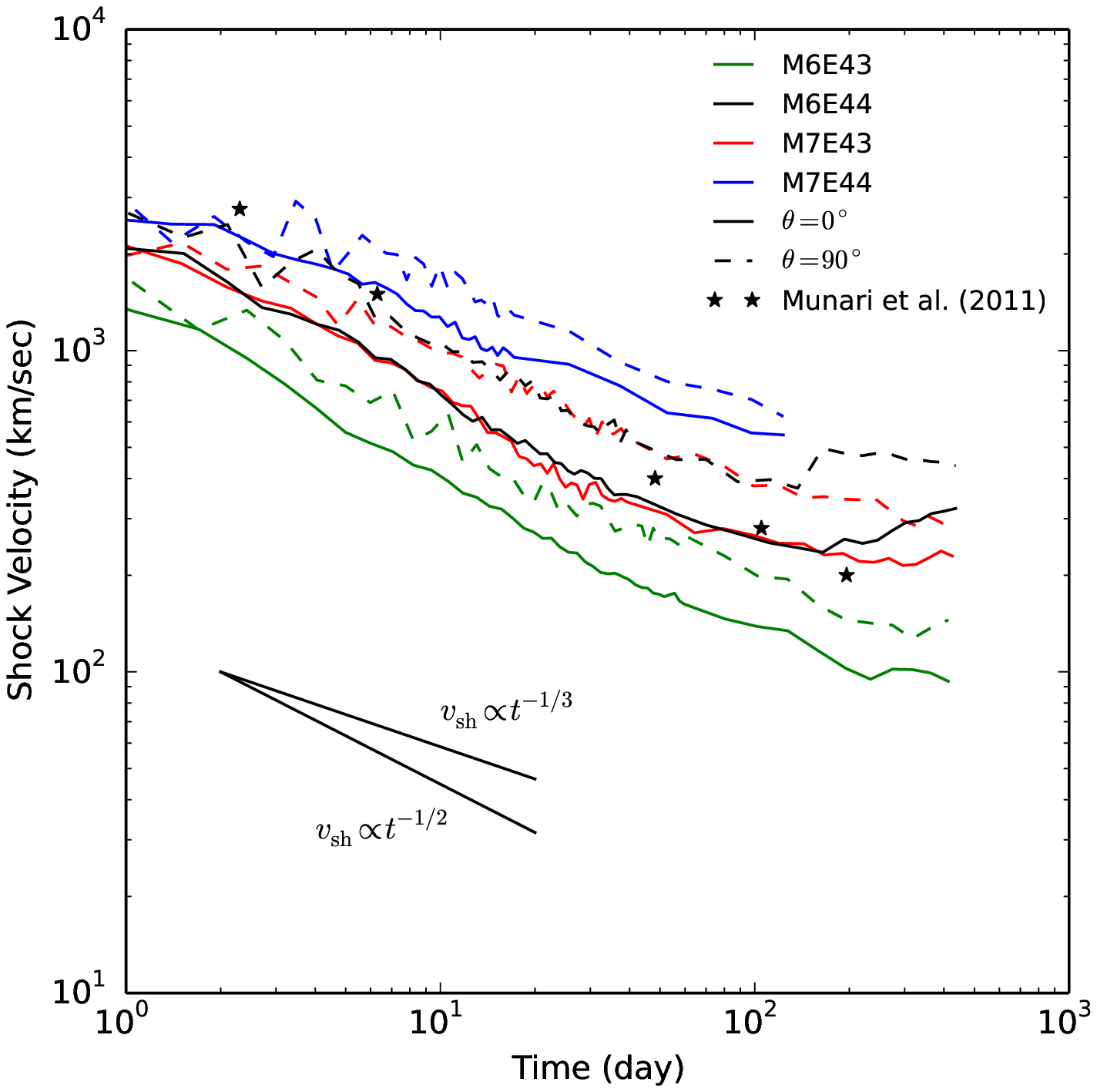}
\end{center}
\caption{\label{fig_shock_time} Time evolution of angle-averaged forward shock radius (left) and velocity (right). 
Different colors represent different eruption simulations in Table~\ref{tab_simulations}.
Solid lines indicate the averaged shock radius (velocity) in the orbital plane. 
Dashed lines show averaged shock radius (velocity) in the poleward region.
Star symbols represent the observed ejecta speed from \cite{2011MNRAS.410L..52M}.
}
\end{figure}

Figure~\ref{fig_shock_time} shows the averaged forward shock evolution in the orbital plane and poleward 
regions of all considered models in Table~\ref{tab_simulations}.  The shock radius evolution can be 
characterized as $r_{\rm sh} \propto t^\alpha$, where $\alpha$ is an evolution stage-dependent constant. 
Depending on the model, $\alpha$ is around $0.5-0.8$ during the first week.  At the beginning, the 
ejecta are nearly spherically symmetric, but the angle-averaged forward shock evolves slightly 
faster in the orbital plane than in the poleward direction due to the orbital motion.  Later on, 
the shock interaction with the RG reduces the angle-averaged shock speed in the orbital plane,
making $\alpha$ decrease a little for $10 < t < 100$~days.  After about $100$ days, the forward shock 
is far beyond the RG, and $\alpha$ increases again. This behavior is consistent with the 
result in \cite{2008A&A...484L...9W}, but for a different time scale due to differences in binary 
parameters.  Furthermore, if we increase the RG mass loss rate (from model M7E44 to M6E44 
or model M7E43 to M6E43) or decrease the eruption energy (model M6E43 to M6E44 or model M7E43 to M7E44),
$\alpha$ becomes lower.

The angle-averaged shock speed decreases greatly due to the sweeping out of the RG wind.
Starting from a value of $v_{\rm ej} \sim 3,000$~km~s$^{-1}$ at the onset of the 
nova eruption, the average shock speed in the radial direction drops to $v_{\rm ej} 
\sim 100-300$~km~s$^{-1}$ after about a year.  The evolution of shock speed can be characterized as 
another power-law relation with an index of $- \frac{1}{3} \sim -\frac{1}{2}$ (Figure~\ref{fig_shock_time}).  
The speed evolution in Figure~\ref{fig_shock_time} is roughly consistent with the spherically symmetric 
simulation of \cite{2012ApJ...761..182M}.  In their simulation, a better treatment of the radiative 
cooling during shock expansion is implemented, but only in 1D.  Our shock speed is also comparable with 
the observed ejecta speed (star symbols in Figure~\ref{fig_shock_time}) of $\sim 2760$~km~s$^{-1}$ on 
day $+2.3$ and $\sim 200$~km~s$^{-1}$ on day $+196$ from the FWHM of the broad component of $H\alpha$
\citep{2011MNRAS.410L..52M}.  
By fitting the simulated ejecta speed with observations, the best-matched models 
are M6E44 and M7E43.  Furthermore, the 2011 April observation 
also shows that the maximum velocity is $\lesssim 300$~km~s$^{-1}$ \citep{2011A&A...527A..98S}, which is 
also consistent with our simulations.  The decrement of nova ejecta speed with the existence of a CDE 
in our simulations is also consistent with the observed ejecta velocity decrement in RS~Oph.

%-------------------------------------------------------------------------------------------------------------
% Conclusions
%--------------------------------------------------------------------------------------------------------------
\section{CONCLUSIONS}

We have investigated the symbiotic recurrent nova V407~Cyg from the quiescent phase to a nova 
eruption via three-dimensional hydrodynamical simulations.  In quiescent-phase simulations, we examined 
two different mass loss rates of the Mira-type RG.  We studied the CDE produced due to 
the wind-WD interaction in V407~Cyg.
It is found that the induced spiral accretion wake in V407~Cyg is more turbulent 
and the CDE is less efficient than found in RS~Oph 
\citep{2008A&A...484L...9W} due to a relatively high wind-to-orbital speed ratio. 
In addition, we observe no significant wind-captured disk around the WD as reported 
by \cite{2008A&A...484L...9W} in RS~Oph, though the angle-averaged radial density 
profile in the equatorial plane is about $20\%$ higher than that in the poleward direction.  

The shock evolution in the eruption phase also is studied.    
It is found that the forward shock location is highly dependent on the inclination angle,
nova eruption energy, and circumstellar density distribution.
The forward and reverse shock interactions with the RG wind and CDE are also crucial 
factors in the overall evolution.   
In addition, the shock radius evolution can be characterized by a power law, 
$r_{\rm sh} \propto t^{\alpha}$, where $\alpha$ is about $0.5- 1.0$, depending on model and evolution stage.
Our model M6E44 ($M_{\rm ej} = 10^{-6}M_\odot$ and $E_{\rm ej} = 1.2 \times 10^{44}$~erg) and 
model M7E43 ($M_{\rm ej} = 10^{-7}M_\odot$ and $E_{\rm ej} = 1.2 \times 10^{43}$~erg) show 
good agreement with the observed ejecta speed from \cite{2011MNRAS.410L..52M}.
The shock evolution in the presence of CDE in our simulations is similar to what has been 
observed  in V407~Cyg and RS~Oph.

Further work in this series of papers will include investigation of thermal 
and non-thermal emission in V407~Cyg and other symbiotic nova systems 
at different wavelengths, in particular in gamma-ray, X-ray, and radio. 
Similar analysis can be applied to other RNe and CNe as well.
 
%-------------------------------------------------------------------------------------------------------------
%  Acknowledgments
%--------------------------------------------------------------------------------------------------------------

\acknowledgments

KCP would like to thank continuous support from F.-K. Thielemann. 
This work was supported by the European Research Council (ERC) grant FISH, 
by PASC High Performance Computing Grant DIAPHANE, 
and by the Theoretical Institute for Advanced Research in Astrophysics (TIARA) 
in the Academia Sinica Institute of Astronomy and Astrophysics (ASIAA).
PMR and RET acknowledge support by the NSF Division of Astronomical Sciences under AAG 14-13367.
FLASH was developed largely by the DOE-supported ASC/Alliances Center for Astrophysical 
Thermonuclear Flashes at the University of Chicago. 
Analysis and visualization of simulation data were completed using the analysis toolkit {\tt yt}
\citep{2011ApJS..192....9T}. 

%--------------------------------------------------------------------------------------------------------------------------

%\bibliography{ref}

%-----------------------------------------------------------------------------------------------
% Table
\clearpage

\begin{deluxetable}{ccccccc}
\tabletypesize{\scriptsize}
\tablecolumns{7}
\tablecaption{Simulation Parameters \label{tab_simulations}}
\tablewidth{0pt}
\tablehead{ \colhead{Abbreviation} \vspace{-0.0cm} & $\dot{M}_{\rm wind}^a$ & $A^b$ & $v_{\rm wind}^c$ & $T_{\rm wind}^d$ & $E_{\rm ej}^e$ & $M_{\rm ej}^f$ \\
 &  $(M_\odot\ {\rm yr}^{-1})$ & (AU) & (km sec$^{-1}$) & (K) & (erg) & $(M_\odot$)
 }
 \startdata
 \vspace{-0.65cm} \\
 \cutinhead{Quiescent phase}
 M6$^{\dagger}$  & $10^{-6}$ & 16 & 20 & 7,000 & --- & --- \\
 M7  & $10^{-7}$ & 16 & 20 & 7,000 & --- & --- \\
 \cutinhead{Eruption phase}
 M6E43  & $10^{-6}$ & 16 & 20 & 7,000 & $1.2 \times 10^{43}$ & $10^{-7}$ \\
 M6E44$^{\dagger}$  & $10^{-6}$ & 16 & 20 & 7,000 & $1.2 \times 10^{44}$ & $10^{-6}$ \\
 M7E43  & $10^{-7}$ & 16 & 20 & 7,000 & $1.2 \times 10^{43}$ & $10^{-7}$ \\
 M7E44  & $10^{-7}$ & 16 & 20 & 7,000 & $1.2 \times 10^{44}$ & $10^{-6}$ \\
\enddata
\tablenotetext{\dag}{Reference simulation}
\tablenotetext{a}{RG mass loss rate }
\tablenotetext{b}{Binary separation}
\tablenotetext{c}{RG wind speed}
\tablenotetext{d}{Effective wind temperature}
\tablenotetext{e}{Eruption energy}
\tablenotetext{f}{Ejecta mass}
\end{deluxetable}

%------------------------------------------------------------------------------------------------
% Figures: 

%------------------------------------------------------------------------------------------------
% Figures: 

%------------------------------------------------------------------------------------------------
% Figure: 

%------------------------------------------------------------------------------------------------
% Figure: 

%------------------------------------------------------------------------------------------------
% Figures: 

\clearpage
%------------------------------------------------------------------------------------------------
% Figures: 

\clearpage

%------------------------------------------------------------------------------------------------
% Figure 2: 
%------------------------------------------------------------------------------------------------
% Figures: 

\clearpage

%------------------------------------------------------------------------------------------------
% Figures: 

%------------------------------------------------------------------------------------------------
% Figures: 
\clearpage

%------------------------------------------------------------------------------------------------
% Figures: 

%------------------------------------------------------------------------------------------------
% Figures: 

%------------------------------------------------------------------------------------------------
% Figures: 

\clearpage

%------------------------------------------------------------------------------------------------
% Figures: 

%------------------------------------------------------------------------------------------------
% Figures: 
\clearpage

%------------------------------------------------------------------------------------------------

\end{document}